\documentclass[a4paper,fleqn,usenatbib]{mnras}
\usepackage[T1]{fontenc}
\usepackage{ae,aecompl}
\usepackage{amsmath,amssymb}
\usepackage{slashed}
\usepackage{ulem}

\usepackage{color}
\usepackage{hyperref}
\hypersetup{colorlinks,citecolor=blue}
\usepackage{amssymb}
\usepackage{xcolor}
\usepackage{graphics,graphicx}
\usepackage{epsf} 
\usepackage{amsmath,amssymb}
\usepackage{longtable}
\usepackage{lastpage}

\def\arcsec{$^{\prime\prime}$}
\def\~{$\sim$}

\def\Lq{\textquotedblleft}
\def\arcsec{$^{\prime\prime}$}
\def\~{$\sim$}

\def\Lq{\textquotedblleft}

\newcommand{\myemail}{yogesh.chandola@pmo.ac.cn}
\title[H{\sc i} maps of mid-infrared bright BCDGs]{GMRT H{\sc i} mapping of mid-infrared bright Blue Compact Dwarf  Galaxies W1016+3754 \& W2326+0608}
\author[Chandola Y., et al.]{Yogesh Chandola$^{1,2,3}$ \thanks{\myemail}, Chao-Wei Tsai$^{3,4,5}$\thanks{ cwtsai@nao.cas.cn}, Di Li$^{3,6,7}$\thanks{ dili@nao.cas.cn}, Chandreyee Sengupta$^{1}$, \newauthor Yin-Zhe Ma$^{7,8,1,9}$\thanks{ma@ukzn.ac.za}, Pei Zuo$^{3,10,11}$ \\
$^{1}$ Purple Mountain Observatory, Chinese Academy of Sciences (CAS), 10, Yuan Hua Road, Qixia District, Nanjing, 210023, China \\
$^{2}$ Inter-University Centre for Astronomy and Astrophysics (IUCAA), Post bag 4, Ganeshkhind, Pune, 411007, India \\	
$^{3}$ National Astronomical Observatories, CAS, 20A, Datun Road, Chaoyang District, Beijing, 100101, China \\
$^{4}$Institute for Frontiers in Astronomy and Astrophysics, Beijing Normal University,  Beijing 102206, China\\
$^{5}$ University of Chinese Academy of Sciences, Beijing 100049, People's Republic of China\\
$^{6}$Research Center for Intelligent Computing Platforms, Zhejiang Laboratory, Hangzhou 311100, China\\
$^{7}$NAOC-UKZN Computational Astrophysics Centre (NUCAC), University of KwaZulu-Natal, Durban, 4000, South Africa \\
$^{8}$School of Chemistry and Physics, University of KwaZulu-Natal, Westville Campus, Durban, 4000, South Africa \\
$^{9}$ National Institute for Theoretical and Computational Sciences (NITheCS), South Africa \\
$^{10}$Kavli Institute for Astronomy and Astrophysics, Peking University, 5 Yiheyuan Road, Beijing 100871, People's Republic of China\\
$^{11}$ International Centre for Radio Astronomy Research (ICRAR), University of Western Australia, 35 Stirling Highway, Crawley,\\ WA 6009, Australia\\
}
\date{\today}
\pubyear{2023}
\begin{document}
	\label{firstpage}
	\pagerange{\pageref{firstpage}--\pageref{lastpage}}
	\maketitle
	\begin{abstract}
		 We present the results from deep 21 cm H{\sc i} mapping of two nearby blue compact dwarf  galaxies (BCDGs), W1016+3754 and W2326+0608, using the  Giant Metrewave Radio Telescope (GMRT).  These BCDGs are bright in mid-infrared (MIR) data and  undergoing active star formation. With the GMRT observations, we  investigate the role of cold neutral gas as the fuel resource of the current intensive star formation activity. Star formation in these galaxies is  likely to be due to the infall of H{\sc i} gas triggered by gravitational perturbation from nearby galaxies. The BCDG W2326+0608 and nearby galaxy SDSS J232603.86+060835.8 share a common H{\sc i} envelope. We find star formation takes place in the  high H{\sc i} column density gas ($\gtrsim 10^{21}$\,cm$^{-2}$) regions for both BCDGs.  The recent starburst and infall of metal-free gas have kept the metallicity low for the BCDG W1016+3754.  The metallicity for W2326+0608 is higher, possibly due to tidal interaction with the nearby galaxy SDSS J232603.86+060835.8.

	\end{abstract}

\section{Introduction}
\label{sec1}  
Blue Compact Dwarf  Galaxies (BCDGs) are compact galaxies with an appearance of extragalactic H{\sc ii} regions \citep{1970ApJ...162L.155S}, and have  different physical properties compared to other dwarf galaxies \citep{1966ApJ...143..192Z,cairos2001ApJS..133..321C,cairos2001ApJS..136..393C, gildepaz2003ApJS..147...29G}. BCDGs are characterized by intense narrow emission lines superposed on a faint blue continuum ($M_{\rm B}$ $>$ $-$18) powered by starburst no older than a few Myr \citep{1999Ap&SS.265..489K}. The metallicities of these sources lie in the range 1/50 Z$_{\odot}$ $<Z<$ 1/2 Z$_{\odot}$ \citep{izotovthuan1999ApJ...511..639I,hunterhoffman1999AJ....117.2789H,lopezsanchez2009A&A...508..615L}. Due to the dust ejected and heated by Type {\sc{II}} supernovae (SNe {\sc{II}}) in compact ($<$ 50 pc), dense star-forming regions, known as super star clusters (SSC), \Lq\textit{active  star-forming }" BCDGs such as SBS 0335$-$052 are brighter in infrared emission  than \Lq\textit{passive star-forming}" BCDGs like {\sc{I}} Zw 18, where star formation is very diffuse in much extended regions ($>$ 100 pc) \citep{2004A&A...421..555H}. In mid-infrared (MIR) surveys such as \textit{Wide-field Infrared Survey Explorer} \citep[\textit{WISE};][]{2010AJ....140.1868W}, due to hot (200-1500 K) and small grain size dust, active BCDGs have redder infrared emission across the \textit{WISE} band (W1-W3: 3.4, 4.6, 12 $\mu$m).

Among the optically selected star-forming dwarf galaxies, those with the reddest MIR colours are found to be the most compact, with blue optical colour ($g-r < -$0.3 mag) and very high specific star formation rate (sSFR $>$ 10$^{-8}$ yr$^{-1}$) \citep{2016ApJ...832..119H, 2017ApJ...847...38Y, 2018arXiv180610149R}. From their optical selection, \cite{2017ApJ...847...38Y} termed such compact ($<$ 1 kpc in size) galaxies with  blue colour, low stellar mass and extreme starburst as \lq\textit{blueberry}' galaxies. These objects represent the fainter and lower mass end of higher redshift ($z>0.2$) green pea \citep[][]{amorin2010ApJ...715L.128A} and Lyman-$\alpha$ Emitters \citep[LAE;][]{gawiser2007ApJ...671..278G}, and are very rare \citep{2011A&A...536L...7I,2014A&A...561A..33I, 2017ApJ...847...38Y}.  From early \textit{WISE} data, \cite{2011ApJ...736L..22G} discovered two such BCDGs, W0801+2640 and W1702+1803, with very red colour (W1[3.4 $\mu$m]$-$W2[4.6 $\mu$m]$>$ 2 mag) and very low metallicities ($<$ 1/8 Z$_{\odot}$). This indicates \textit{WISE} colour may help in searching low metallicity \textit{active star-forming} BCDGs. However, MIR bright BCDGs are found to have a wide range in their metallicities \citep[][]{2011A&A...536L...7I,2014A&A...561A..33I}. The metal-poor nature of some active star-forming BCDGs is still not clearly understood. It is believed either the starburst has taken place very recently and/or there is an infall of metal-poor gas which has diluted the metal abundances in these sources \citep{2018MNRAS.477..392L}. 

H{\sc i} content, morphology, distribution and kinematics of metal-poor BCDGs have been studied in literature to understand the reasons for triggering star-burst and low metallicities \citep{2007A&A...464..859P,2009MNRAS.397..963E,2010MNRAS.403..295E,lopezsanchez2010A&A...521A..63L,lopezsanchez2012MNRAS.419.1051L,2013A&A...558A..18F, 2014A&A...566A..71L, 2016MNRAS.463.4268T}. However, there are only a few H{\sc i} studies towards MIR bright actively star-forming (sSFR $> 10^{-8}$ yr$^{-1}$) BCDGs due to their rarity \cite[e.g. SBS0355-052,][]{2009MNRAS.397..963E}. We have investigated the H{\sc i} contents and conditions in the extreme starburst and low-intermediate metallicity environment using a sample of 11 nearby BCDGs with {\it WISE} red colour and bright MIR emission (Chandola et al. submitted). With Arecibo Observatory (AO), we performed deep H{\sc i} observations towards these sources, including W1017+3754 and W2326+0608 with bright H{\sc i} emission,  F$_{\rm peak}$ (H{\sc i}) $>$ 5 mJy,  which were observed further with the Giant Metrewave Radio Telescope (GMRT) to understand the H{\sc i} kinematics and the environment of these BCDGs.

In this paper, we present results from the deep 21 cm H{\sc i} observations of W1016+3754 and W2326+0608, two nearby,  mid-infrared bright, starburst BCDGs using the GMRT. Despite the facts of being MIR bright objects with active star formation, sharing similar stellar mass  and having larger nearby galaxies, these two BCDGs differ in their metallicities by nearly an order of magnitude. In Sec.~\ref{sec2}, we describe the properties of these two sources and associated multiwavelength data used in this paper (also see Table~\ref{sourcechar}). 
The details of GMRT observations and data reduction procedure are provided in Sec.~\ref{sec3}. H{\sc i} maps and results are shown in Sec.~\ref{sec4}. We discuss the results from the study on the H{\sc i} contents and environments of these two galaxies in  Sec.~\ref{sec5} and summarize in Sec.~\ref{sec6}. Throughout this paper, we assume a concordance cosmology with $H_{0}=70\,{\rm km}\,{\rm s}^{-1}\,{\rm Mpc}^{-1}$, $\Omega_{\rm m}=0.3$ and $\Omega_\Lambda=0.7$.
We adopt a solar metallicity (Z$_{\sun}$) of $12+\log$(O/H)$=8.69$~\citep{2009ARA&A..47..481A}. 
Magnitudes are reported in the Vega system. 
\begin {table}
\caption {Properties of the two BCDGs}
\begin {center}
{\scriptsize
\begin {tabular}{c c c }
\hline
Name    & W1016+3754 & W2326+0608 
\\
\hline
Right Ascension$^{a}$ &10:16:24.5 &  23:26:03.6
\\
(J2000)&& \\
Declination$^{a}$ &+37:54:45.8& +06:08:15 
\\    
(J2000)&&\\
Redshift & 0.00388$\pm$0.00001$^{a}$ & 0.01678$\pm$0.00006$^{b}$
\\
$D_\mathrm{L}$ [Mpc]$^{c}$& 16.7 &   72.8
\\ 
scale [pc/arcsec]$^{c}$ &80 & 341
\\ 
12 + log (O/H) & 7.57$\pm$0.01$^{d}$ &   8.39$\pm$0.03$^{b}$
\\   
log (M$_{*}$/M$_{\odot}$)$^{e}$ & 7.2 & 7.0
\\
SFR [M$_{\odot}$/yr]$^{e}$ & 0.04 & 0.3
\\
log (sSFR)  [yr$^{-1}$]$^{e}$ & $-$8.6 & $-$7.5
\\
W1[3.4 $\mu$m]$-$W2[4.6 $\mu$m]$^{f}$  & 0.55$\pm$0.06 & 1.31$\pm$0.18\\ 
$\rm[mag]$  &  & \\
W2[4.6 $\mu$m]$-$W3[12 $\mu$m]$^{f}$  & 3.94$\pm$0.10 & 5.65$\pm$0.15\\
$\rm[mag]$  &     & \\
W4[22 $\mu$m]$^{f}$  & 6.70$\pm$0.07 & 6.57$\pm$0.08\\
$\rm[mag]$ &     & \\
\hline     
\end {tabular}
}
Notes: 
$D_{\rm L}$ stands for luminosity distance and SFR  for Star Formation Rate. \\     
References:
$^{a}$SDSS~\citep{2015ApJS..219...12A},$^{b}$~Zhang Ludan et al. in preparation, $^{c}$based on cosmological model mentioned in Sec.~\ref{sec1}, $^{d}$~\cite{2012MNRAS.427.1229I}, $^{e}$~Chandola et al. submitted,
$^{f}$AllWISE~\citep{2013yCat.2328....0C}

\end {center}
\label{sourcechar}
\end {table}
\begin{figure*}
	\centering
	
	\hspace{-0.6cm}	
	\includegraphics[scale=0.4]{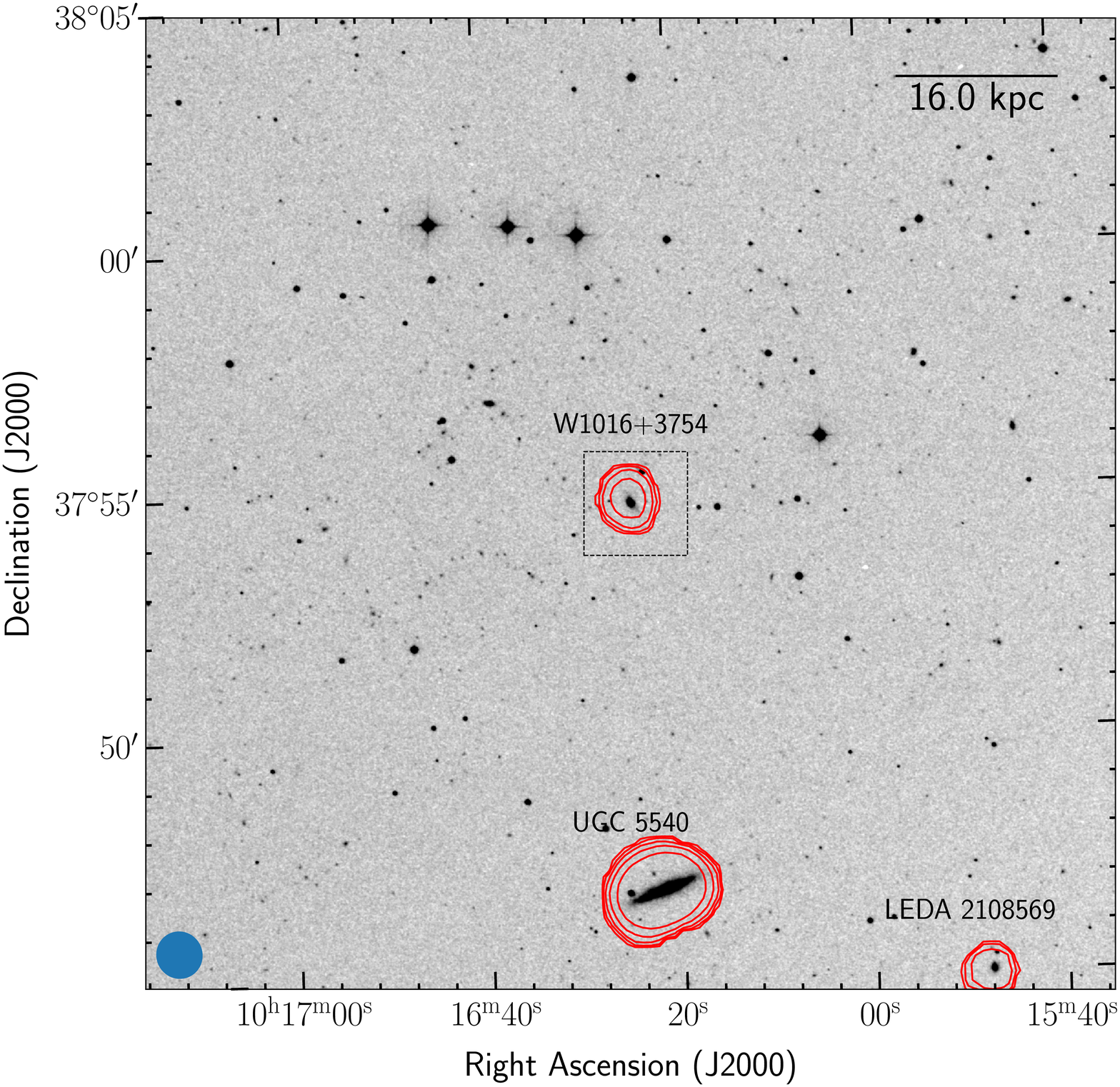}
	\vspace{-0.6cm}
	\hbox{
		\includegraphics[scale=0.17]{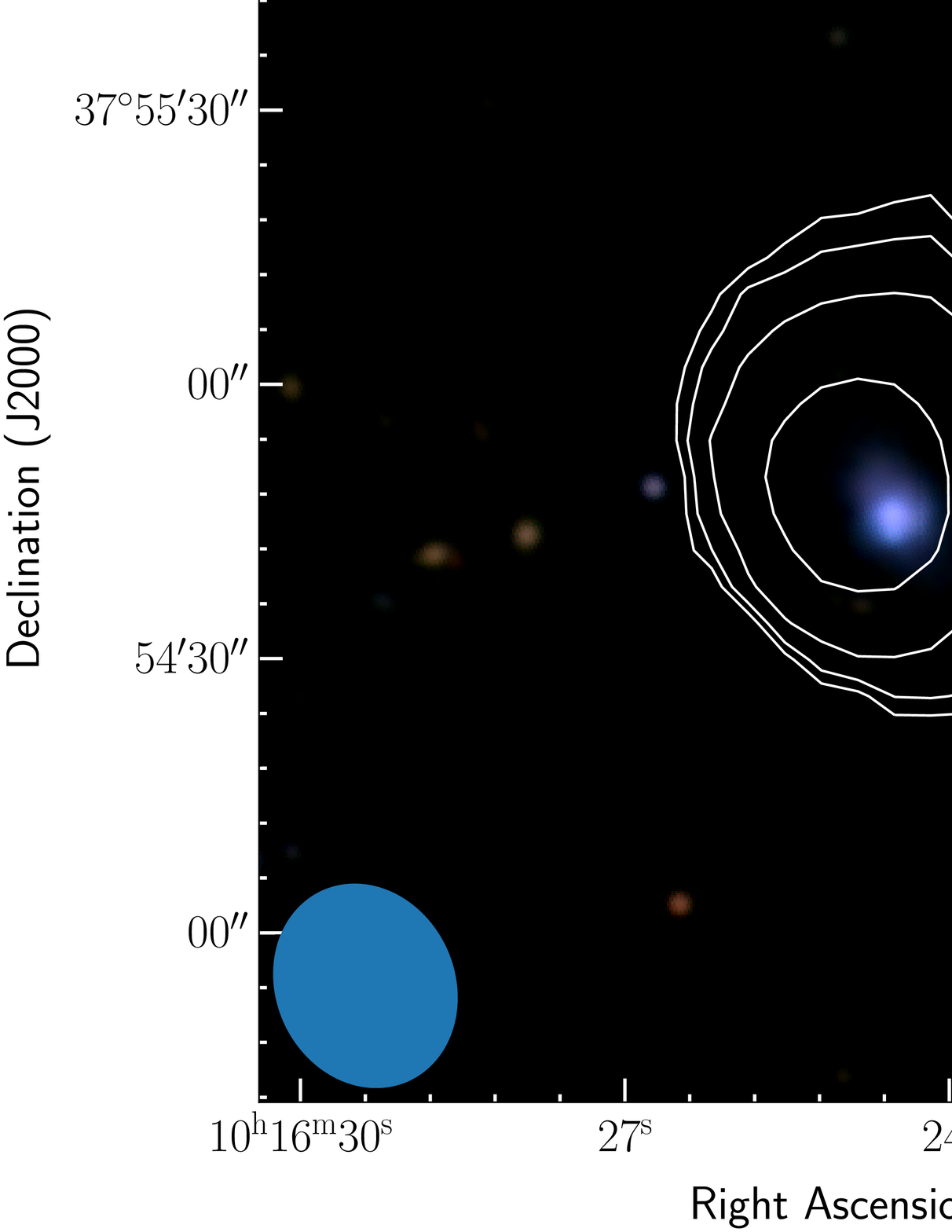} 
		\hspace{-0.5cm}
		\includegraphics[scale=0.17]{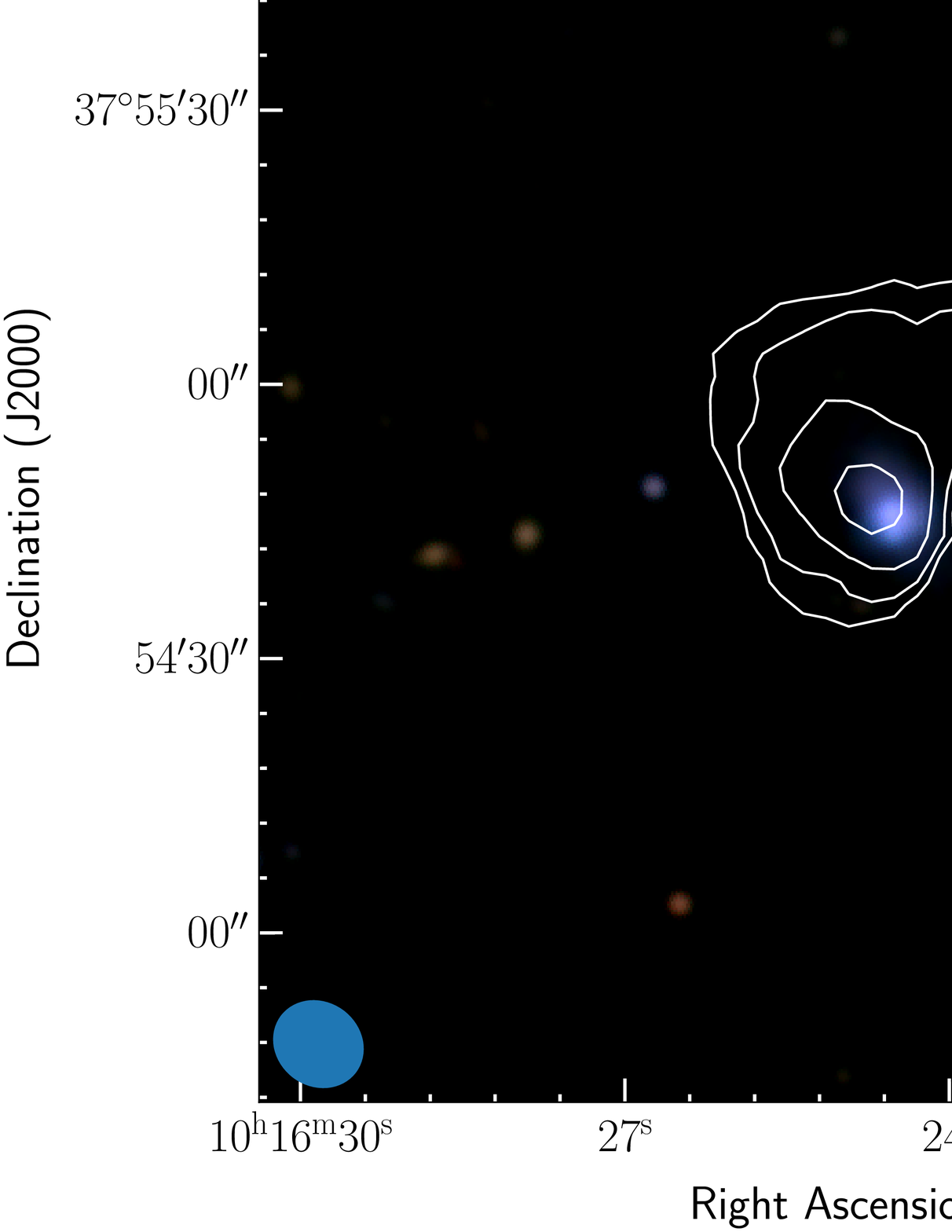}
	}
	\vspace{0.3cm}
	
	\caption{\textbf{Top panel}: H{\sc i} emission line integrated flux density (moment zero) contours for W1016+3754 field overlaid on  Digitized Sky Survey (DSS) r-band optical image. The contours are at  (1, 2, 4, 8, 16) $\times$ 36 mJy beam$^{-1}$ km s$^{-1}$  which correspond to H{\sc i} column densities (1, 2, 4, 8,16) $\times$ 1.2 $\times$  10$^{19}$ cm$^{-2}$. Synthesized beam size is 58.7\arcsec $\times$ 57.0\arcsec, P.A. 20.4$^{\circ}$ (shown in blue at bottom left corner of the image).  The first contour in all H{\sc i} maps (including this map) in different figures in this paper represent the 3-$\sigma$ H{\sc i} detection limits  given in Table~\ref{HIcub} assuming a minimum of 3 spectral channels are used in the moment map.  \textbf{Bottom left panel}: 
		A close up H{\sc i} map of W1016+3754 galaxy (marked with dashed square in top panel) at spatial resolution 22.9\arcsec$\times$19.6\arcsec, P.A. 24.3$^{\circ}$. Contours are at (1, 2,4,8) $\times$ 25 mJy beam$^{-1}$ km s$^{-1}$   which correspond to H{\sc i} column densities (1, 2, 4, 8) $\times$ 6.2 $\times$ 10$^{19}$ cm$^{-2}$. The overlaid image is from Sloan Digital Sky Survey (SDSS) i-band(red), r-band(green) and g-band (blue). 
		\textbf{Bottom right panel}: Same as in bottom left panel but in higher resolution 10.3\arcsec$\times$9.2\arcsec, P.A. 56.3$^{\circ}$. Contours are at (1, 2, 4, 8) $\times$ 18.0 mJy beam$^{-1}$ km s$^{-1}$  which correspond to H{\sc i} column densities (1, 2, 4, 8) $\times$ 2.1 $\times$ 10$^{20}$ cm$^{-2}$.	
	}
	\label{fig1}
\end{figure*} 

\begin{figure}
	\centering
	
	\hbox{
		
		\includegraphics[scale=0.23]{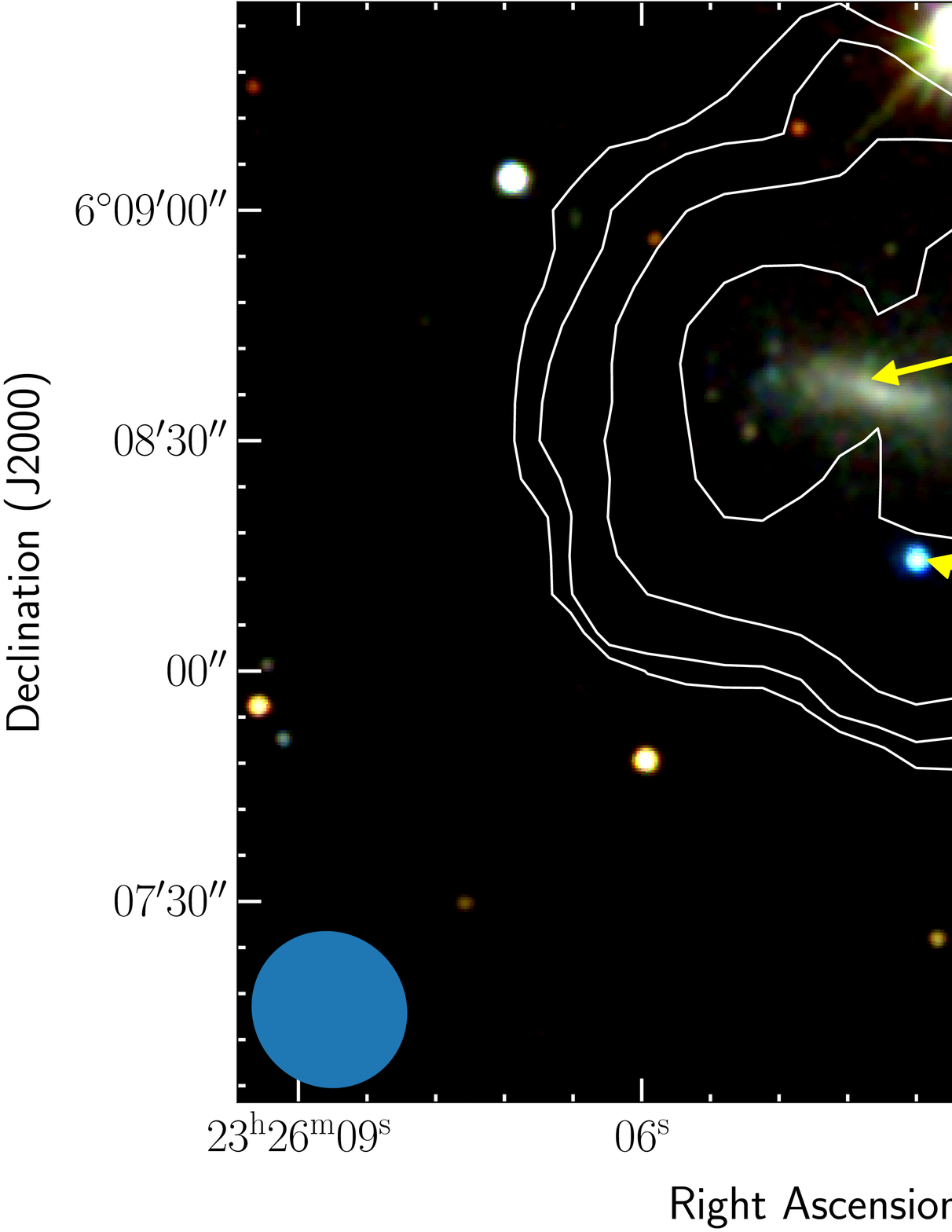}
		
	} 
	
	\caption{H{\sc i} integrated flux density (moment zero) contours from GMRT observations for W2326+0608 overlaid on SDSS optical 3 color image (red: i-band, green: r-band , blue: g-band). The contours are at ( 1, 2, 4, 8) $\times$ 22.2 mJy beam$^{-1}$ km s$^{-1}$  which correspond to H{\sc i} column densities ( 1, 2, 4, 8) $\times$ 5.9 $\times$ 10$^{19}$ cm$^{-2}$.  Synthesized beam size is 20.8\arcsec $\times$ 19.9 \arcsec, P.A. 39.0$^{\circ}$ (shown in blue at left bottom corner of the image).	 		
	}
	\label{fig1a}
\end{figure} 

\section{Properties of the two BCDGs}
\label{sec2}
\subsection{W1016+3754}
W1016+3754 is a blue compact dwarf galaxy with  cometary tadpole-like morphology in optical 
images (Fig.~\ref{fig1}) at redshift $\sim$ 0.0039 which corresponds to a luminosity distance $D_{\rm L}$ $\sim$ 17 Mpc. An edge-on disk galaxy, UGC 5540, is at same redshift, $z=0.0039$,  and angular distance of $\sim 8 \arcmin$ towards the south. Red {\it WISE} MIR colours  ( W1[3.4 $\mu$m]$-$W2[4.6 $\mu$m]$=$ 0.55 mag, W2[4.6 $\mu$m]$-$W3[12 $\mu$m]$=$3.94 mag) and bright 22 $\mu$m emission (W4[22 $\mu$m]$=$ 6.7 mag)~\citep{2013yCat.2328....0C}, in W1016+3754 indicate the presence of hot dust heated by intense star formation. Stellar mass of $\log(M_{\ast}/{\rm M}_{\odot}) \sim 7.2$ 
is estimated using the correlation between optical \textit{ugriz} colours and mass-to-light ratio from \cite{bell2003ApJS..149..289B}. The $K$-corrections were done using the best fit model from \cite{assef2010ApJ...713..970A} to UV-near infrared spectral energy distribution (SED)(Chandola et al. submitted). Using the method of \cite{hao2011ApJ...741..124H} and \cite{kennicutt2012ARA&A..50..531K}, star formation rate (SFR) of $\sim$0.04 ${\rm M}_{\odot}\,{\rm yr}^{-1}$ is estimated from \textit{Galaxy Evolution Explorer} \citep[\textit{GALEX};][]{bianchi2017ApJS..230...24B} Far UV luminosity  corrected for dust extinction with 22$\mu$m values from \textit{WISE}  (Chandola et al. submitted). This gives a specific star formation rate $\sim 10^{-8.6}\,{\rm yr}^{-1}$ for this galaxy. The H$\alpha$ image shows presence of two star formation regions \citep{2016MNRAS.462...92J}. The  equivalent width (EQW) of H$\beta$ line for this system is 96.7{\AA} \citep{2012MNRAS.427.1229I}, implying starburst age of $\sim$2 Myr  \citep{1986A&A...156..111C, 1996ApJS..107..661S}. The oxygen abundance is estimated as, 12+log(O/H) $=$ 7.57 $\pm$ 0.01, $<$ 10 $\%$ of solar metallicity \citep{2012MNRAS.427.1229I}. Some high-ionization lines, such as [Ne{ \sc v}]$\lambda$3426 and [He{ \sc ii}]$\lambda$4686 lines are also detected \citep{2012MNRAS.427.1229I}.  These lines can arise from active galactic nuclei (AGN) \citep{2008ApJ...687..133I}, X-ray binaries \citep{1986Natur.322..511P}, and fast radiative shocks from Supernovae (SNe) or young starburst \citep{1996ApJS..102..161D}, but the lack of X-ray detection in this galaxy \citep{2013ApJ...769...92P} suggests a starburst nature.  [He{ \sc ii}]$\lambda$4686 line could be also attributed to the presence of Wolf-Rayet stars \citep[][]{schaerer1999A&AS..136...35S,crowther2006A&A...449..711C,lopezsanchez2010A&A...516A.104L}.
\begin {table*}
\caption {Details of GMRT H{\sc i} observations}
\begin {center}
\begin {tabular}{ c c c c c c}
\hline
Source        & Date       & Central &  Total    & Flux density/  &   Gain/    \\
&            & Frequency & time$^{\dagger}$ & bandpass &  phase                     \\
&            & [MHz]    &  [hrs]           &  calibrator(s)   & calibrator(s)                      \\
\hline
W1016+3754& 2014 May 24, 25    & 1414.89 & 12 &   3C48, 3C286 & J1035+5628         \\
& 2015 May 9,10  & 1414.89 & 17 &   3C48, 3C147, 3C286  &  J1035+5628                  \\
W2326+0608& 2014 May 24,27 & 1397.07 & 14 &   3C48, 3C286  &  J0022+0014, J2212+0152 \\
& 2015 May 9,10  & 1397.07 & 19 &   3C48, 3C286  &  J0022+0014, J2212+0152                      \\  
\hline
\end {tabular}
\end {center}
\label{obslog2}
$^{\dagger}$:  Including calibration and other overheads.
\end {table*}
\begin {table*}
\caption {Characteristics of H{\sc i} 
	emission cubes.}
\begin {center}
\begin {tabular}{ c c c c c c c c c}
\hline
Source   & \textit{uv} taper    & \textit{R}    & Synthesized &       Vel. res.$^{\ast}$&Pixel& $\Delta S_{\rm rms}$(1 $\sigma$)  & Sensitivity$^{\dagger}$ &Detectable    \\
&            &           &  beam size& $\delta v$&size&[mJy beam$^{-1}$]  &[mJy beam$^{-1}$]&N(H{\sc i})$^{\dagger}$   \\
& [k$\lambda$]&& [\arcsec $\times$ \arcsec, P.A.] & [ km s$^{-1}$]  &[\arcsec] & [channel$^{-1}$] & [km s$^{-1}$]&[10$^{18}$ cm$^{-2}$]     \\
\hline
&&&&&&&&\\
W1016+3754     & 50    & 0 &  3.5\arcsec$\times$ 3.2\arcsec, 72.4$^{\circ}$ & 6.9&1.0 &0.3 & 6.2& 607 \\

&  15   & 0 &  10.3\arcsec$\times$9.2\arcsec, 52.7$^{\circ}$ &6.9 & 2.5& 0.5  & 10.4& 120.6\\
&  8    & 0 &  22.9\arcsec$\times$19.6\arcsec, 24.3$^{\circ}$ &6.9 & 4& 0.7  & 14.5& 35.7\\

&  2    & 5  & 58.7\arcsec$\times$57.0\arcsec, 20.4$^{\circ}$ & 6.9 & 10 & 1.0  & 20.7& 6.8 \\
&&&&&&&&\\
W2326+0608		& 40    & 0 & 4.8\arcsec $\times$ 4.4 \arcsec,81.5$^{\circ}$ & 7.1 & 1.25& 0.4    & 8.52 & 445.7 \\

&  8    & 0 & 20.8\arcsec$\times$ 19.9\arcsec, 39.0$^{\circ}$ & 7.1 & 5   &  0.6    & 12.8 & 34.1   \\  

\hline
\end {tabular}
\end {center}
\textit{R}:Robustness parameter,
$^{\ast}$: Velocity resolution,
$^{\dagger}$:  At 3$\sigma$ level for single channel. 
\label{HIcub}
\end {table*}
\begin{table*}
	\caption {Parameters derived from  global H{\sc i} profiles in W1016+37 and W2326+0608 field}
	\begin{center}          
		\begin{tabular}{ccccccccc}
			\hline
			Sl. no. & Field &Source  &H{\sc i} velocity & FWHM & $S_{\rm peak}$ &$\Delta\rm S_{\rm rms}$ &Integrated flux & $\log (M_\mathrm{HI}/{\rm M}_{\odot})$ \\
			&       &   &[km s$^{-1}$] & [km s$^{-1}$] & [mJy] &[mJy channel$^{-1}$] &[Jy km s$^{-1}$] &  \\
			\hline
			1 &W1016+3754&W1016+3754       & 1173$\pm$5   &44$\pm$9 & 8.6$\pm$1.0  &0.9 &0.39$\pm$0.06 & 7.4  \\
			2 &W1016+3754&UGC 5540          & 1163$\pm$1   &101$\pm$2 & 53.5$\pm$3.3  &2.0 &4.70$\pm$0.20 & 8.5 \\
			3 &W1016+3754&LEDA 2108569     & 1279$\pm$6   &57$\pm$11 & 8.5$\pm$1.5  &1.4 &0.45$\pm$0.10 & 7.6 \\
			4 &W2326+0608& all & 4991$\pm$2   &97$\pm$4              & 20.3 $\pm$2.0  &1.8 & 1.63$\pm$0.17 & 9.3 \\	
		
			5 &W2326+0608&W2326+0608  & 5023$\pm$7  & 57$\pm$15& 1.5$\pm$0.4 & 0.4 &0.05$\pm$0.03 & 7.8 \\
			
			\hline
		\end{tabular}
	\end{center} 
	\label{sourcecharHI}
\end{table*}
\begin{table}
	\caption{Gaussian fit parameters for W1016+3754 H{\sc i} profile}
	\begin{center}
		\begin{tabular}{cccc}
			\hline
			Component	&Peak flux & Velocity & FWHM \\
			&[mJy]	 & [km s$^{-1}$] & [km s$^{-1}$] \\
			\hline
			1	& 10.3$\pm$0.9 & 1170$\pm$4 & 41$\pm$7 \\
			2	& 4.0$\pm$1.1 & 1213$\pm$9 & 35$\pm$20 \\
			\hline
		\end{tabular}
	\end{center}
	\label{gaussfit}	
\end{table}

\begin{figure*}
	\centering
	\includegraphics[scale=0.38]{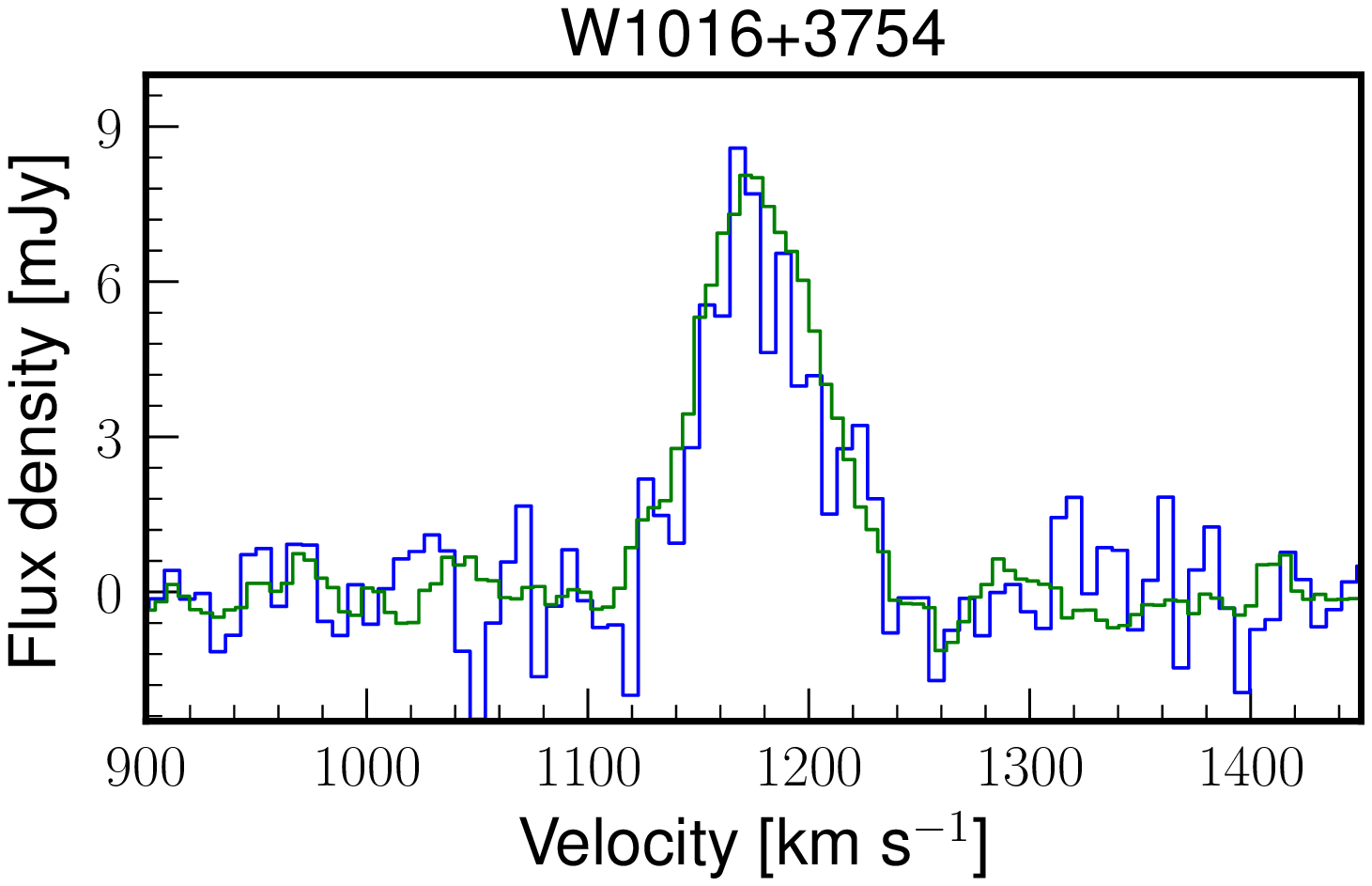}
	\includegraphics[scale=0.38]{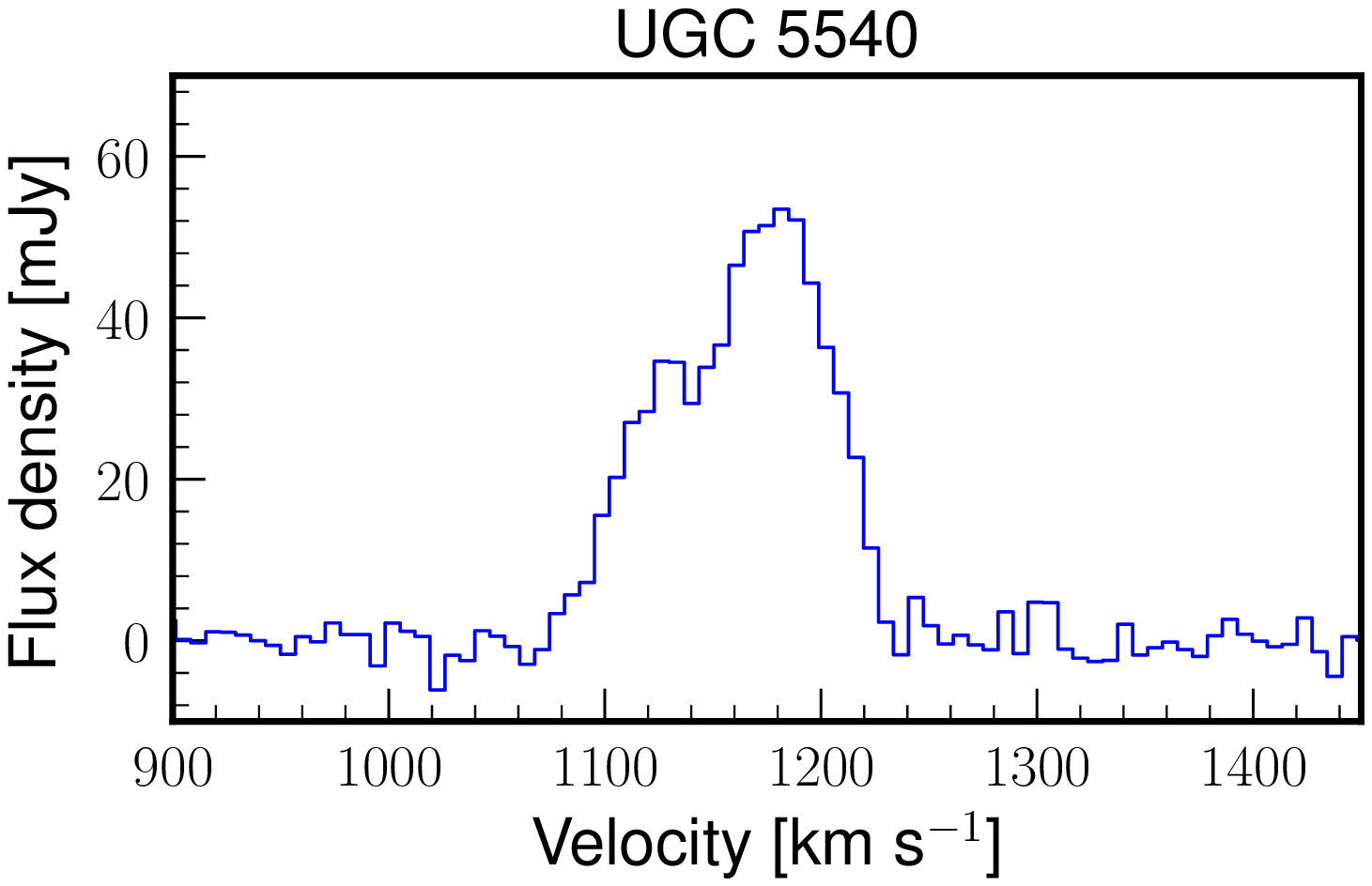}
	\includegraphics[scale=0.38]{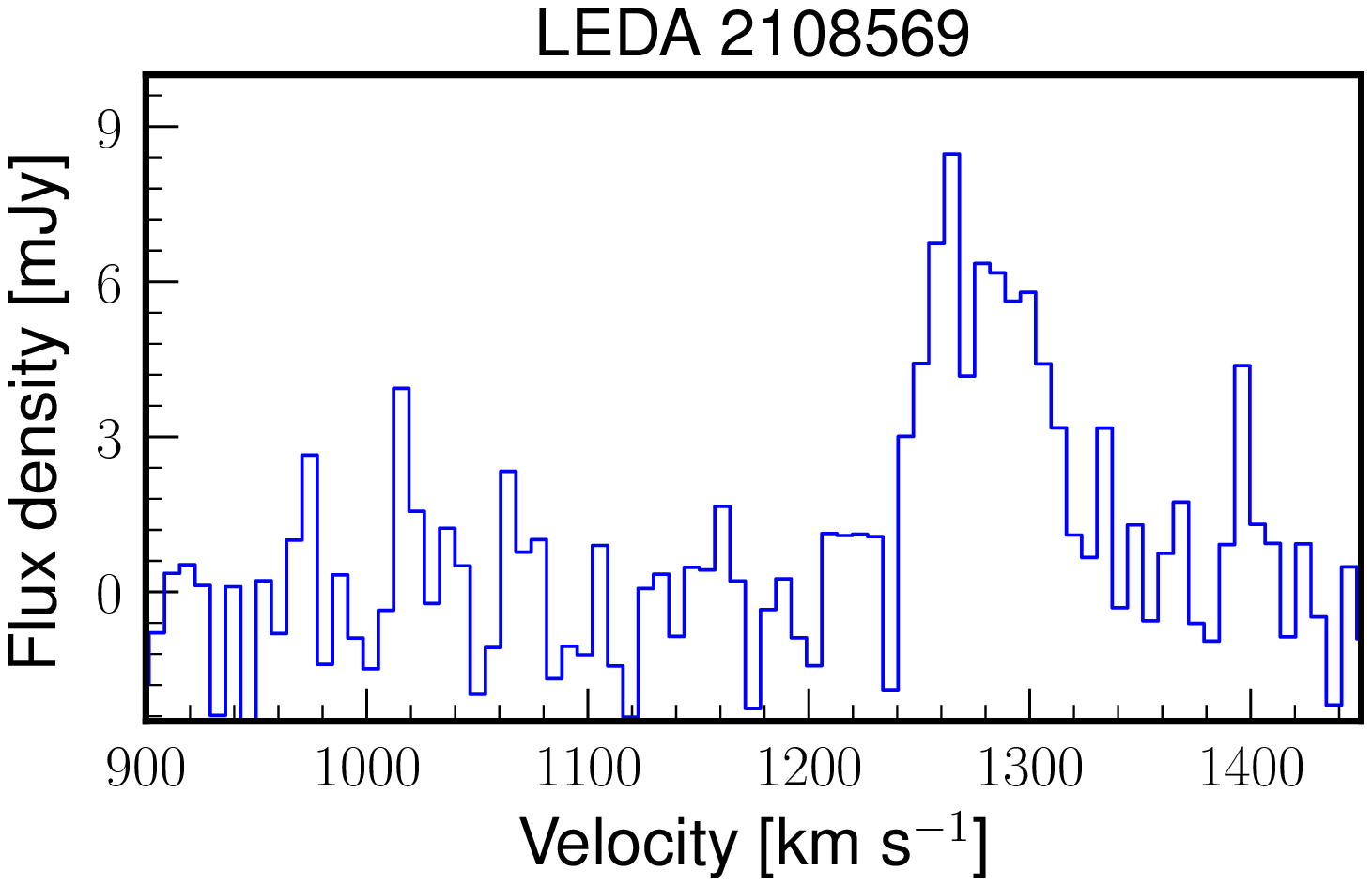}
	\caption{Integrated H{\sc i} line profiles of W1016+3754, UGC 5540, and south-west dwarf galaxy LEDA 2108569. For W1016+3754,  the profile from the Arecibo observation (Chandola et al. submitted) is also shown in green for comparison.}
	\label{fig2}
\end{figure*}     
\begin{figure*}
	\centering
	\hbox{
		\includegraphics[scale=0.23]{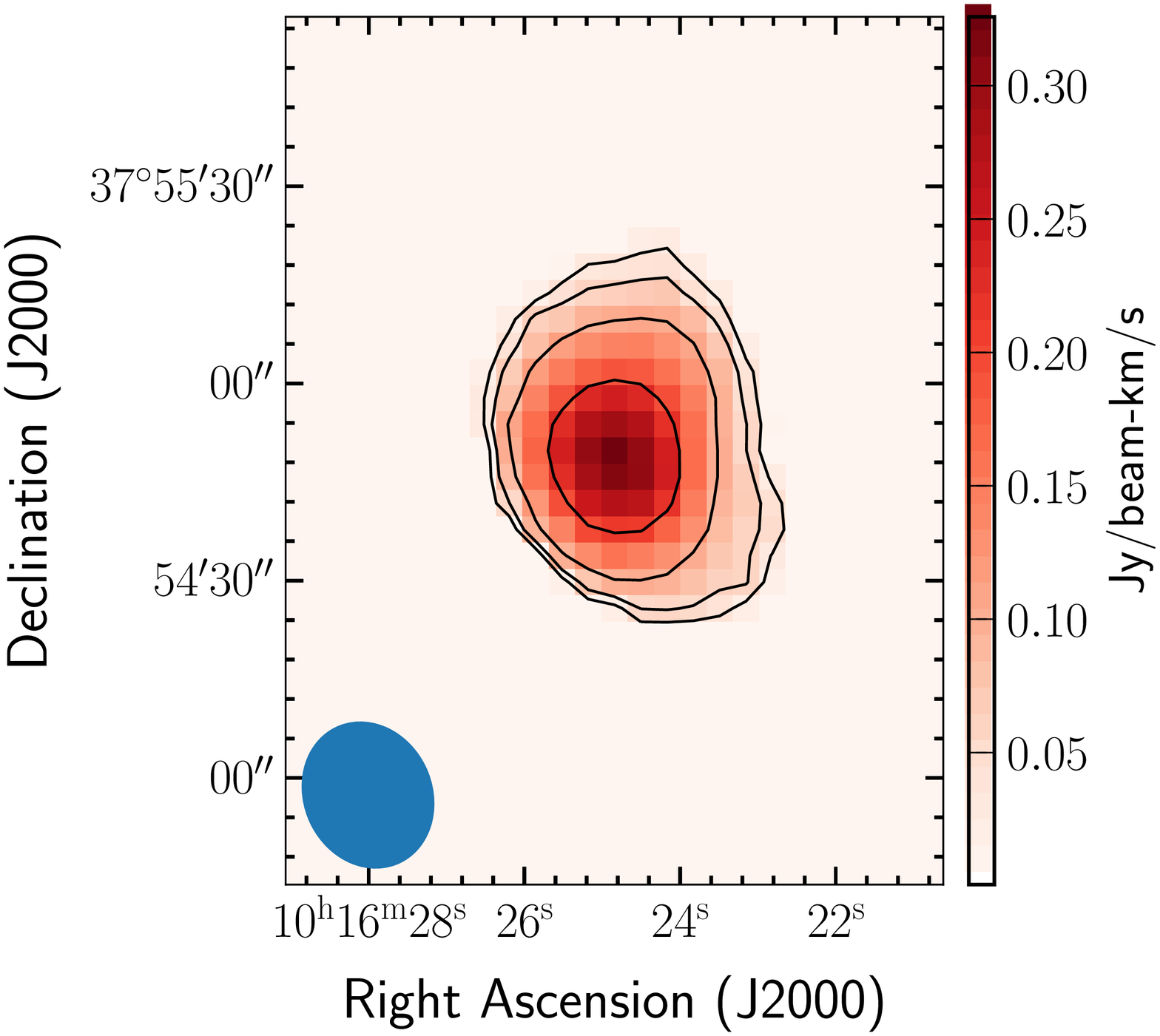}
		\hspace{0.33cm}
		\includegraphics[scale=0.23]{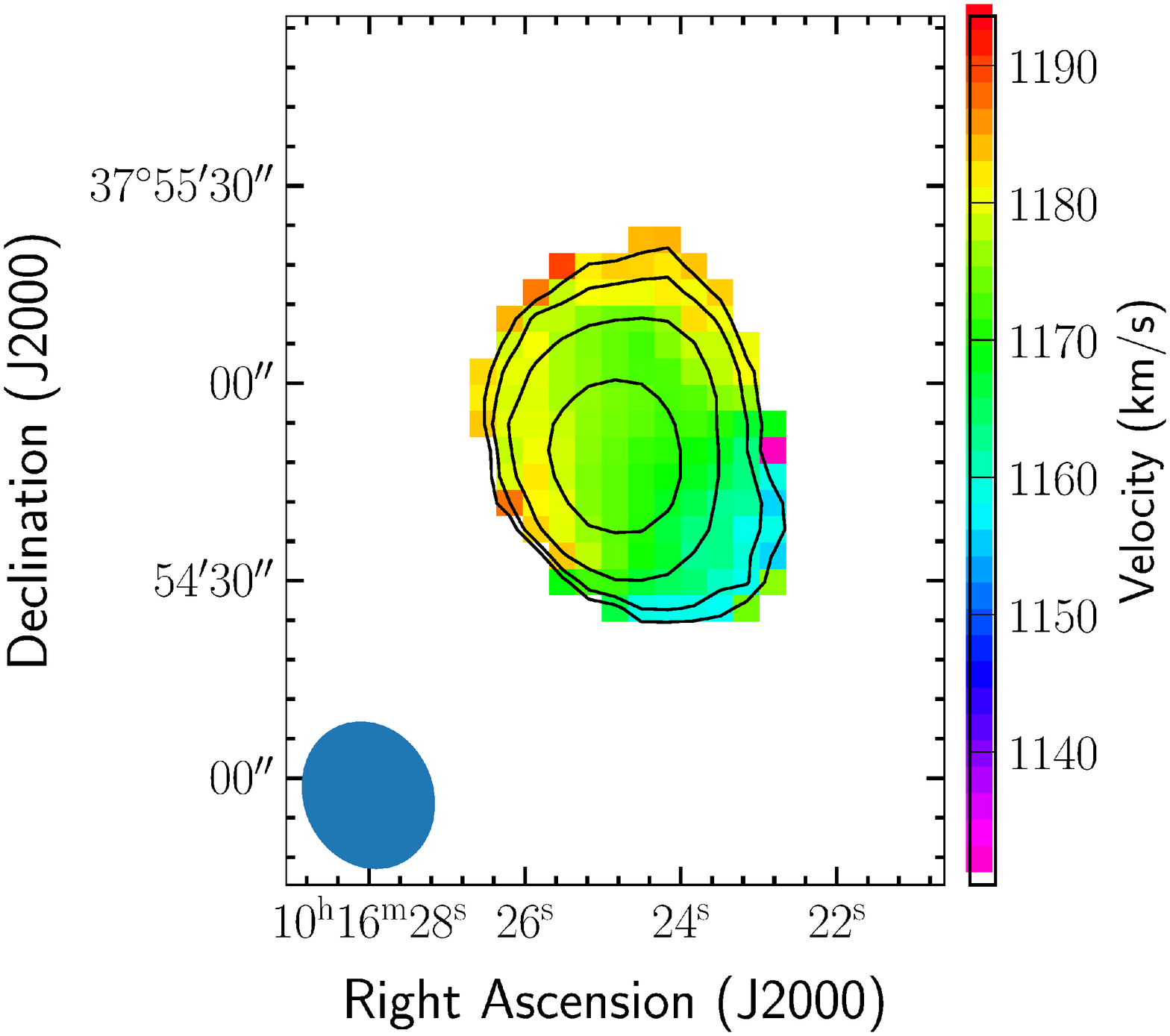}
		\hspace{0.33cm}
		\includegraphics[scale=0.23]{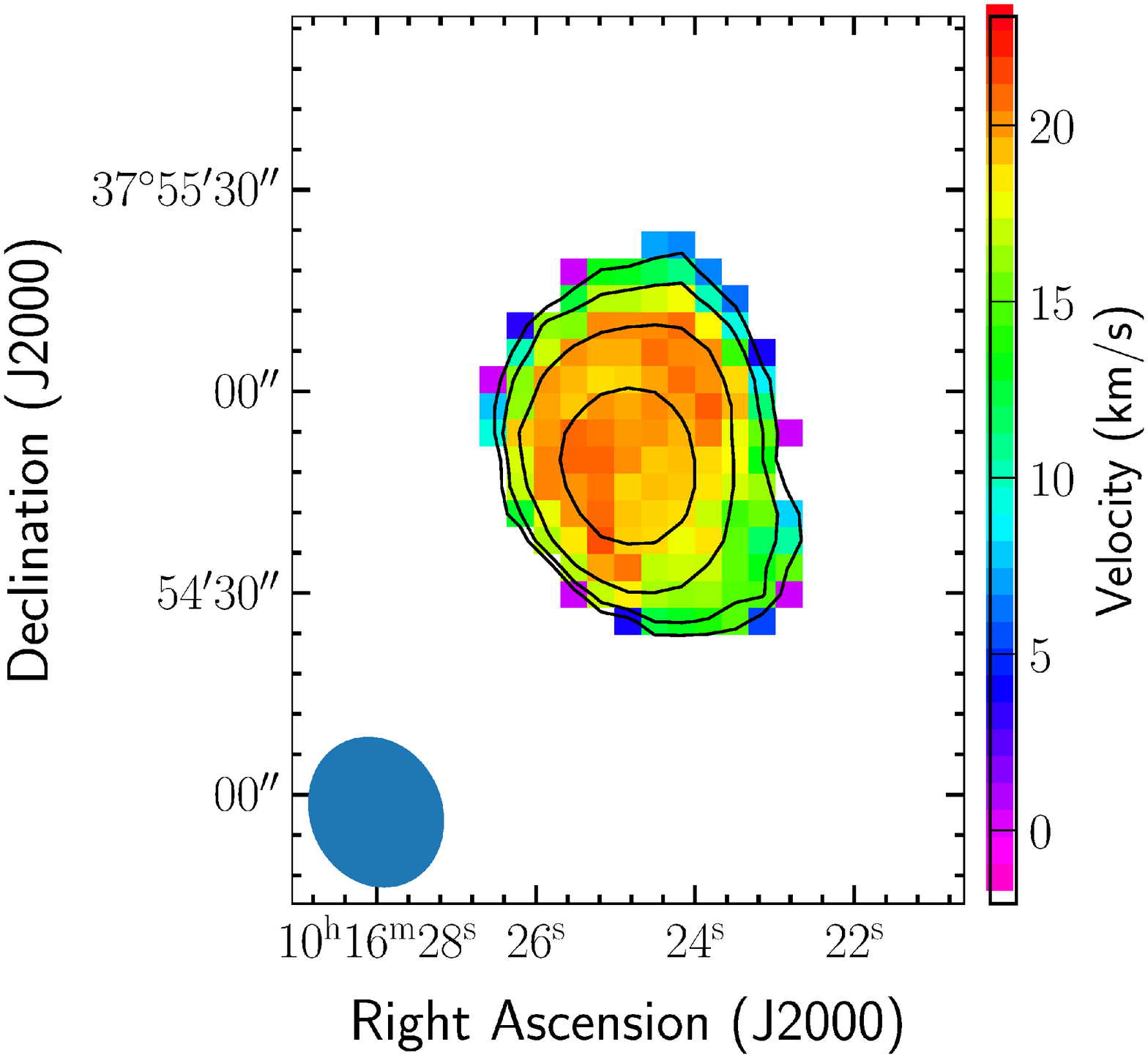}
	} 
	\hbox{
		\includegraphics[scale=0.23]{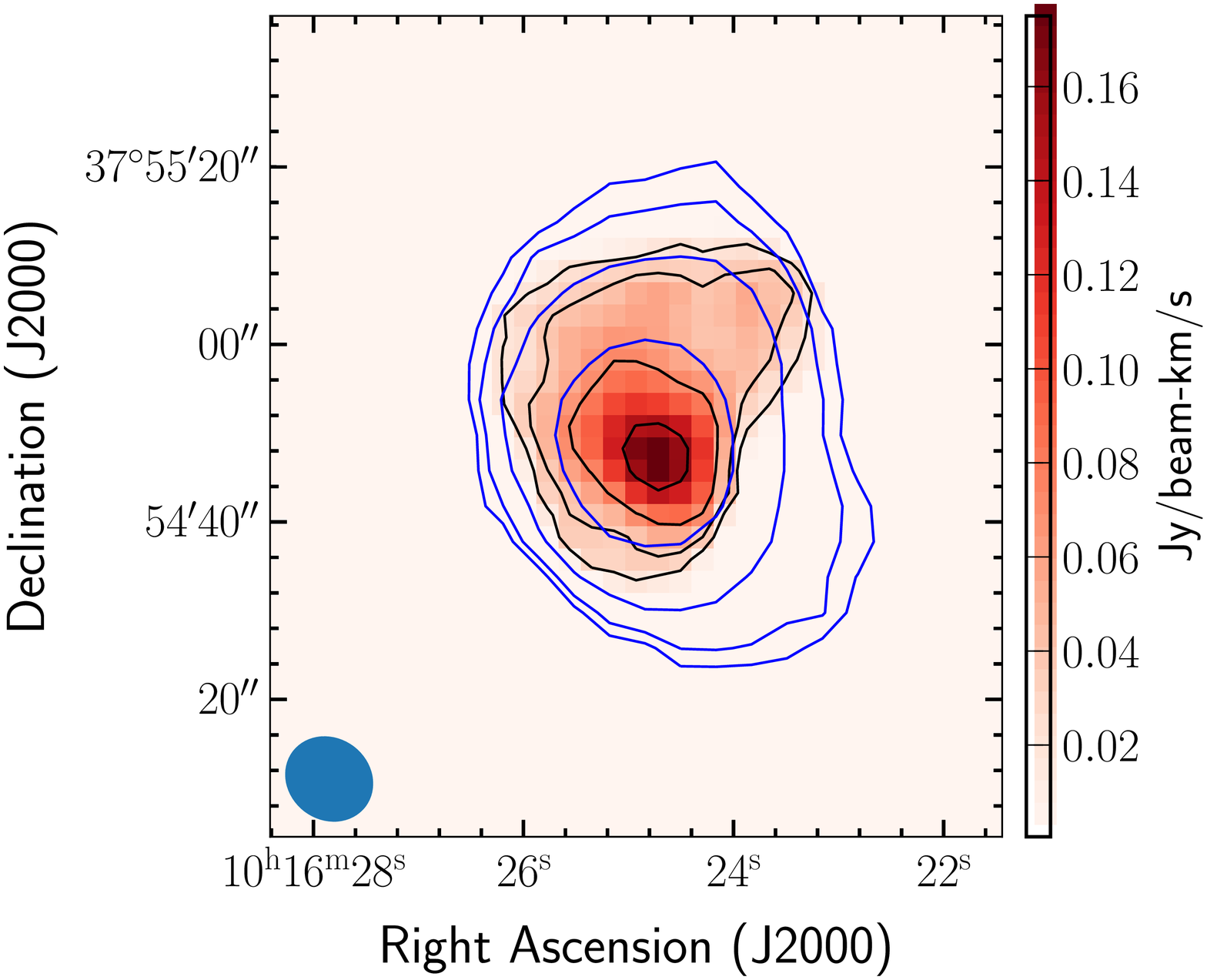}
		\includegraphics[scale=0.23]{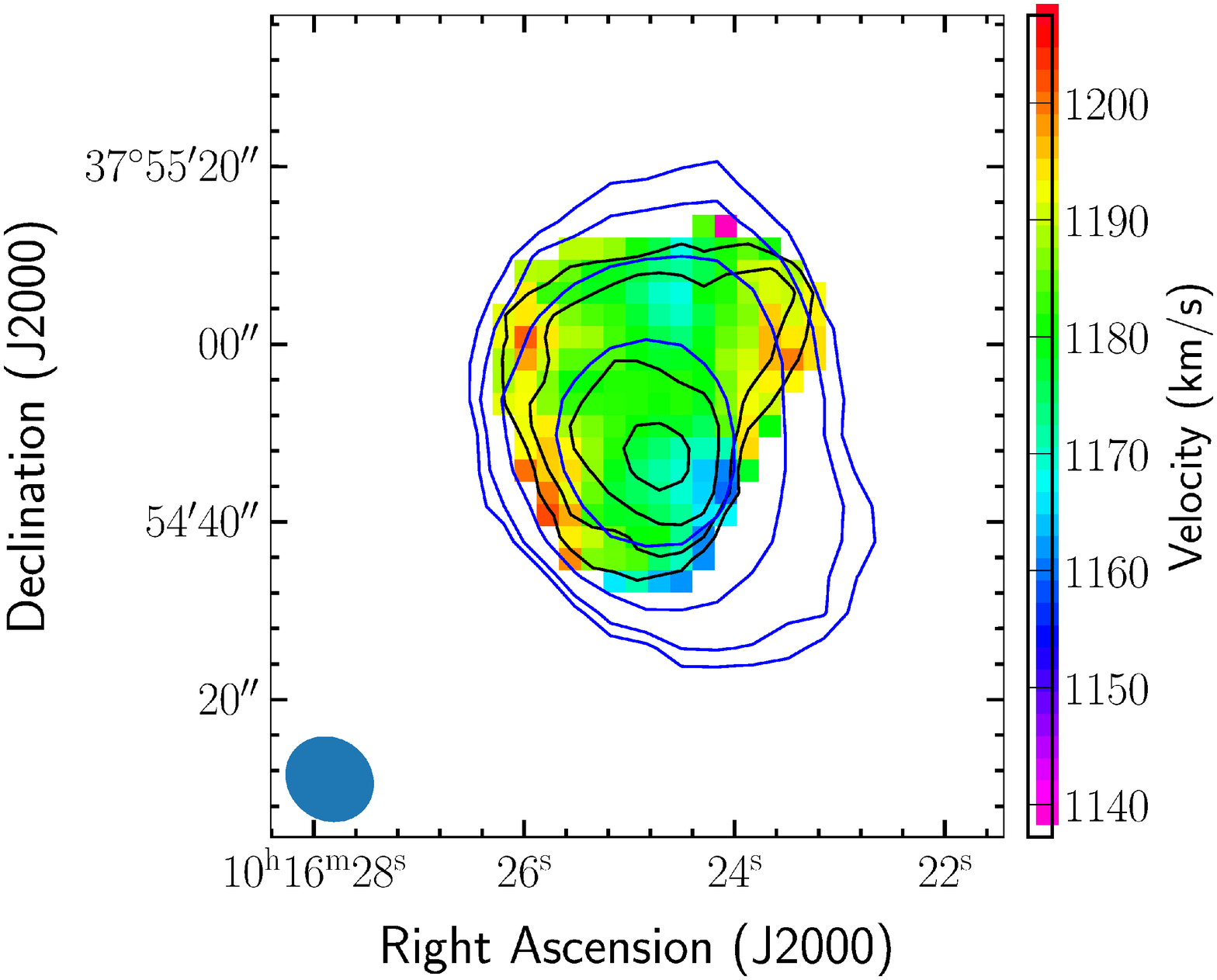}
		\includegraphics[scale=0.23]{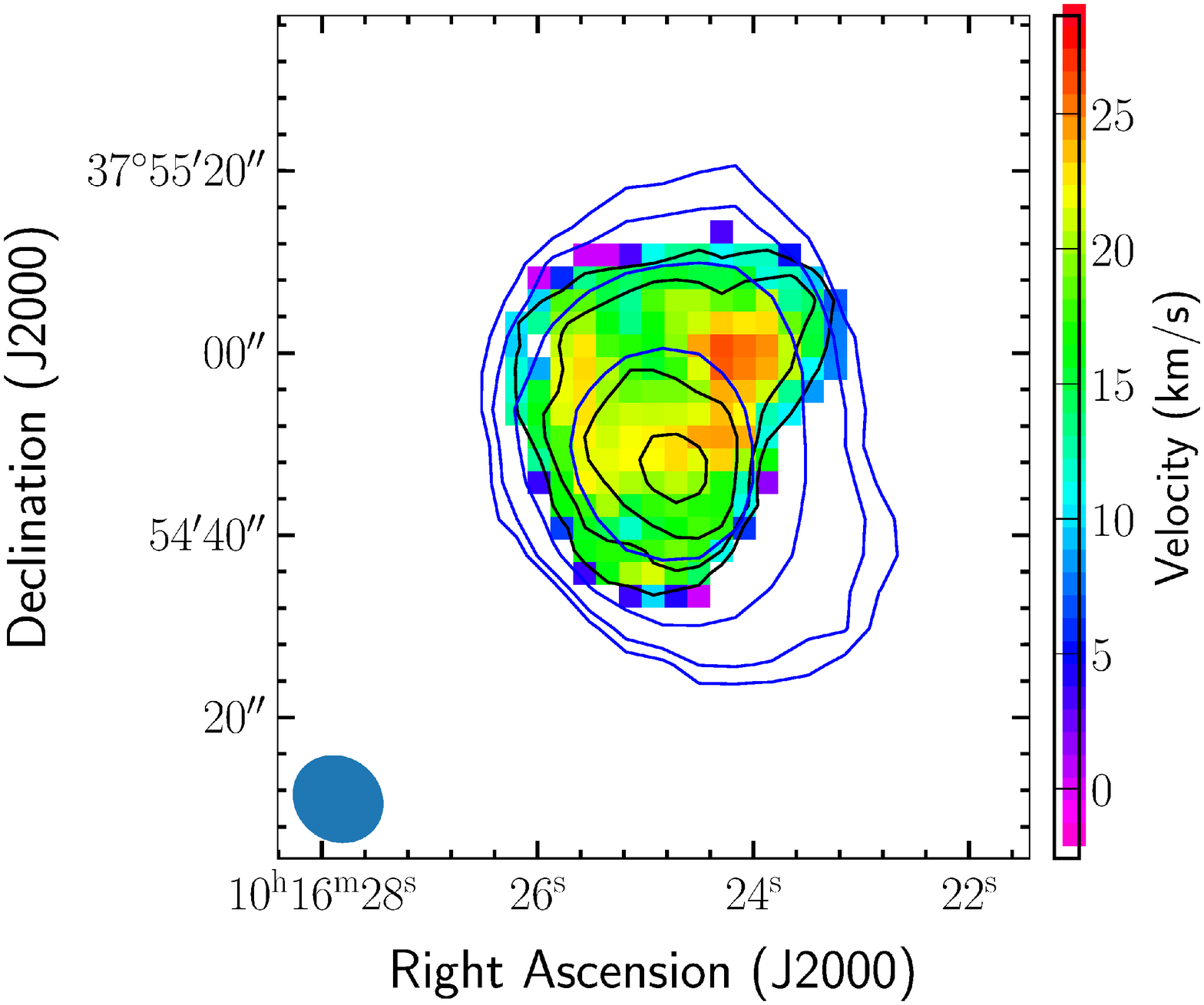}
	}   		
	
	\caption{\textbf{Top panel:} Intermediate resolution H{\sc i} moment zero contours for the BCDG W1016+37 overlaid on colour scale moment zero, moment one and moment two maps. 
		Moment maps are at spatial resolution 22.9\arcsec$\times$19.6\arcsec, P.A. 24.3$^{\circ}$. Contours are at (1, 2, 4, 8) $\times$ 25 mJy beam$^{-1}$ km s$^{-1}$  which correspond to H{\sc i} column densities (1, 2, 4, 8) $\times$ 6.2$\times$  10$^{19}$ cm$^{-2}$.
	\textbf{Bottom panel:} Higher resolution H{\sc i} moment zero contours for the BCDG W1016+37 overlaid on colour scale moment zero, moment one and moment two maps. 
	Moment maps are at spatial resolution 10.3\arcsec$\times$9.2\arcsec, P.A. 52.7$^{\circ}$. Contours are at (1, 2, 4, 8) $\times$ 18.0 mJy beam$^{-1}$ km s$^{-1}$  which correspond to H{\sc i} column densities (1, 2, 4, 8) $\times$ 2.1$\times$  10$^{20}$ cm$^{-2}$. For comparison, intermediate resolution moment zero contours in blue colour are also shown.}             
	\label{fig3}
\end{figure*}

\begin{figure}
	\centering
	\includegraphics[scale=0.5]{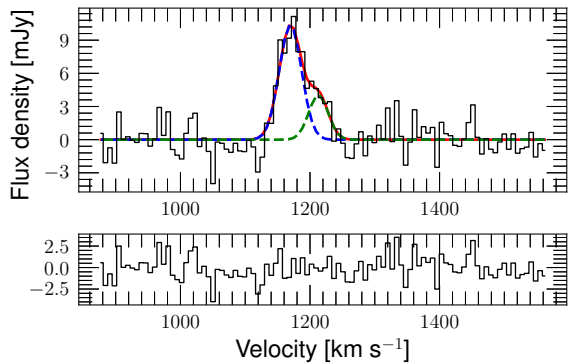}
	\caption{Integrated H{\sc i}  line profile for W1016+3754 extracted from 10.3 $\times$ 9.2\arcsec, P.A. 52.7$^{\circ}$ resolution cube. It shows two Gaussian components with dashed lines in green and blue. The combined Gaussian profile is shown in red colour. Residuals from the fit are shown in the lower panel.}
	\label{fig5}
\end{figure}	

\begin{figure}
	\centering
	\includegraphics[scale=0.31]{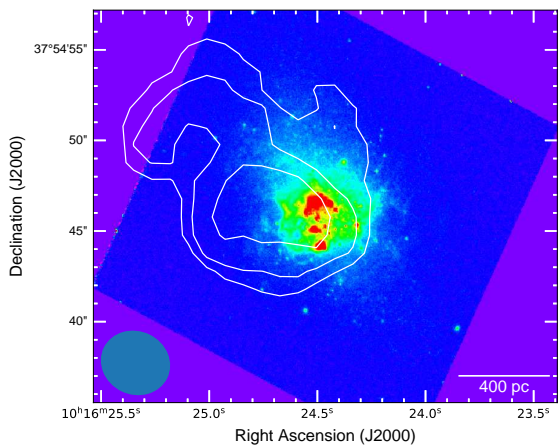}
	\caption{ H{\sc i} total intensity contours  for W1016+3754  overlaid on \textit{Hubble Space Telescope (HST) WFC3 F606W} image. The contours are at (1, 2, 4) $\times$ 10.8 mJy beam$^{-1}$ km s$^{-1}$ which correspond to  H{\sc i} column densities  (1, 2, 4) $\times$ 10.7 $\times$10$^{20}$cm$^{-2}$ for a synthesized beam of 3.5\arcsec $\times$  3.2\arcsec, P.A. 72.4$^{\circ}$.}
	\label{fig6}
\end{figure}

\begin{figure}
	\centering
	
	\hbox{
	
		\includegraphics[scale=0.55]{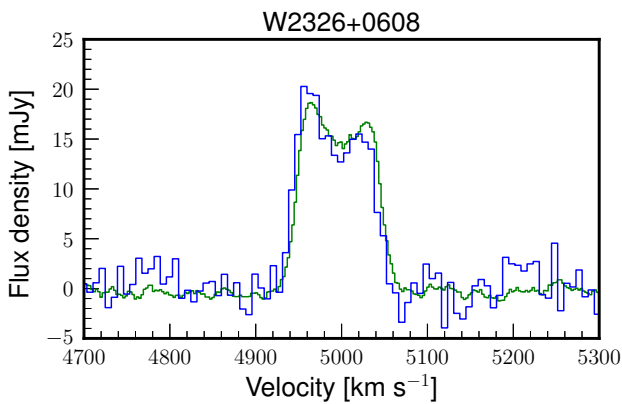}
	} 
	
	\caption{ Integrated H{\sc i} line profile for W2326+0608 from the GMRT observation in blue and the Arecibo observation in green (Chandola et al. submitted) .	 		
	}
	\label{fig7}
\end{figure} 

\begin{figure*}
	\centering
	\vbox{
		\hbox{
					
			\includegraphics[scale=0.25]{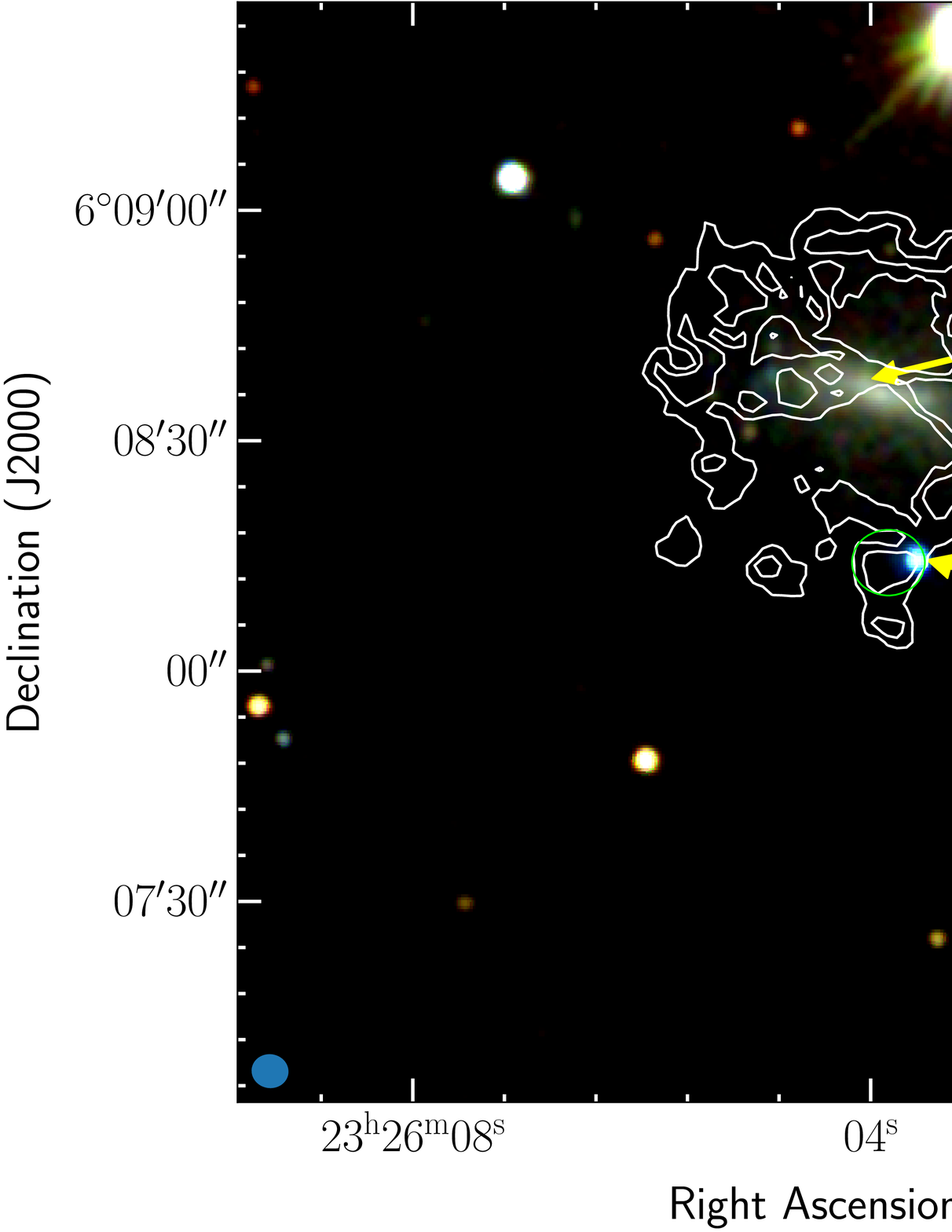}
			
			\includegraphics[scale=0.55]{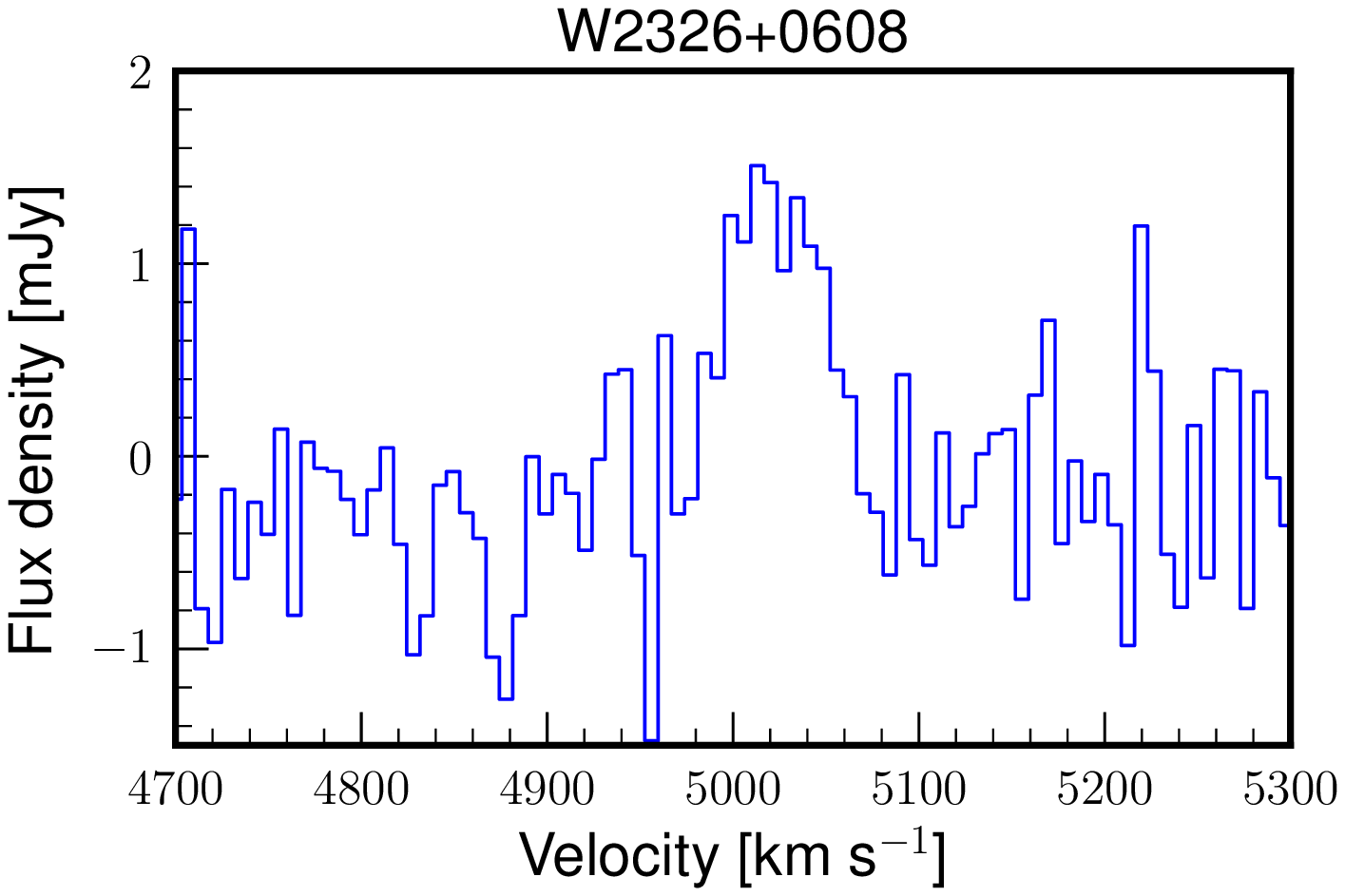}	
		} 
		
	}    				
	\caption{\textbf{Left panel:} H{\sc i} integrated flux density contours for W2326+0608 overlaid on SDSS 3 colour image (red: i-band, green: r-band, blue: g-band). The contours are at  (1, 2, 4) $\times$ 14.8 mJy beam$^{-1}$ km s$^{-1}$  which correspond to  H{\sc i} column densities (1, 2, 4) $\times$ 7.7 $\times$ 10$^{20}$cm$^{-2}$ for a synthesized beam of 4.8\arcsec $\times$  4.4\arcsec, P.A. 81.5$^{\circ}$.  H{\sc i} high density region  close to the BCDG is marked with green circle.
		\textbf{Right panel:} Integrated H{\sc i} line profile extracted from the H{\sc i} high density region close to the BCDG W2326+0608.		
	}
	\label{fig8}
\end{figure*}

\subsection{W2326+0608}
W2326+0608 is a BCDG at $z=0.01678$ ($D_{\rm L} \sim 73\,{\rm Mpc}$). A nearby galaxy, SDSS J232603.86+060835.8, is also seen towards the north at an angular  distance of $\sim$ 22.7 \arcsec (see Fig.~\ref{fig1a}). This angular distance corresponds to a projected distance of $\sim$ 8 kpc if this system is at same redshift as the BCDG. The stellar mass is estimated as $\log(M_{\ast}/{\rm M}_{\odot}) \sim 7.0$ (Chandola et al. submitted). The red MIR colour, W2[4.6 $\mu$m]$-$W3[12 $\mu$m] $=$ 5.65 mag and rather strong 22 $\mu$m emission, W4[22 $\mu$m] $=$ 6.6 mag, compared to the optical emission suggest strong dust emission. The SFR is estimated to be $\sim$ 0.3 M$_{\odot}$ yr$^{-1}$  using FUV luminosity from \cite{assef2010ApJ...713..970A} model SED and corrected for dust extinction with 22 $\mu$m emission values from \textit{WISE} (Chandola et al. submitted). The oxygen abundance estimate is 8.39$\pm$0.03 ($\sim$ 1/2 Z$_{\odot}$) using the method of \cite{2006A&A...448..955I} (Zhang Ludan et al. in preparation). 

\section{GMRT Observations and data reduction}
\label{sec3}
W1016+3754 and W2326+0608 were observed with the GMRT full array and total intensity mode in May 2014 and May 2015.   A base band bandwidth of 16 MHz with 512 channels used during observations provided a velocity resolution ($\delta v$) of $\sim$ 7 km s$^{-1}$. The observational details are provided in  Table~\ref{obslog2}. Flux and bandpass calibrators were observed for $\sim$15 minutes after every 3-4 hours of intervals while phase calibrators were observed for $\sim$ 5 minutes after every $\sim$ 46 minutes of observation on the target source. Total GMRT observing time  for W1016+3754 and W2326+0608, including calibration and other overheads, was 29 and 33 hours respectively.

The GMRT data were reduced using the NRAO Astronomical Image Processing System (\textsc{AIPS})  software\footnote{\url{http://www.aips.nrao.edu/index.shtml}}. Initially, before any calibration the bad data due to dead antennas, radio frequency interference (RFI) and other problematic issues were flagged. Then the antenna gain solutions were calculated using task \textsc{CALIB} and bandpass solutions were determined using \textsc{BPASS}. The data were split by applying the bandpass and gain solutions, and then data from different runs were combined using \textsc{DBCON}. Continuum images were made from line-free channels. Further after a few rounds of self-calibration, the continuum was subtracted from the split \textit{uv} data  using tasks \textsc{UVSUB} and \textsc{UVLIN}. The frequency axis was converted to the H{\sc i} line velocity in the Heliocentric frame of rest with the task \textsc{CVEL}. 
The H{\sc i} data cubes of different resolutions were produced by the task \textsc{IMAGR} using different \textit{uv} taper and robustness parameters.
 The root-mean-square ({\it rms}) noise ($\sigma$) for line-free channels varies from $\sim$0.3 mJy beam$^{-1}$ channel$^{-1}$ in the highest resolution cube to $\sim$1 mJy beam$^{-1}$ channel$^{-1}$ in the lowest resolution cube. The details of H{\sc i} cubes such as synthesized beam sizes corresponding to different \textit{uv} constraints, velocity resolutions, pixel sizes, {\it rms} noises in line free channels  including the  H{\sc i} detection limit of these cubes are provided in Table~\ref{HIcub}.

  We used the smooth and clip algorithm in the Source Finding Application   \citep[\textsc{SoFiA};][]{2015MNRAS.448.1922S}  to  create H{\sc i} detection masks.  This algorithm first smoothed H{\sc i} cubes using Gaussian kernels of radius 1, 3 and 6 pixels in spatial coordinates, and boxcars of width 1, 3, 7 and 15 channels (see Table~\ref{HIcub} for pixel sizes and velocity resolution per channel).  A threshold of above 5 $\sigma$ was used at each step of smoothing to find detection voxels. After that, detections were merged using a radius of 1 pixel and 1 channel. Only sources with sizes greater than 6 pixels and 3 channels were considered as genuine H{\sc i} detections. H{\sc i} masks generated by this process were used to create moment maps from original H{\sc i} cubes. After applying the primary beam correction, we extracted H{\sc i} profiles from cubes using the Common Astronomy Software Applications \citep[\textsc{CASA};][]{2007ASPC..376..127M} task \textsc{IMVIEW}.

\section{Results}
\label{sec4}
\begin{figure*}
	\centering
	\hbox{ 				
		\includegraphics[scale=0.23]{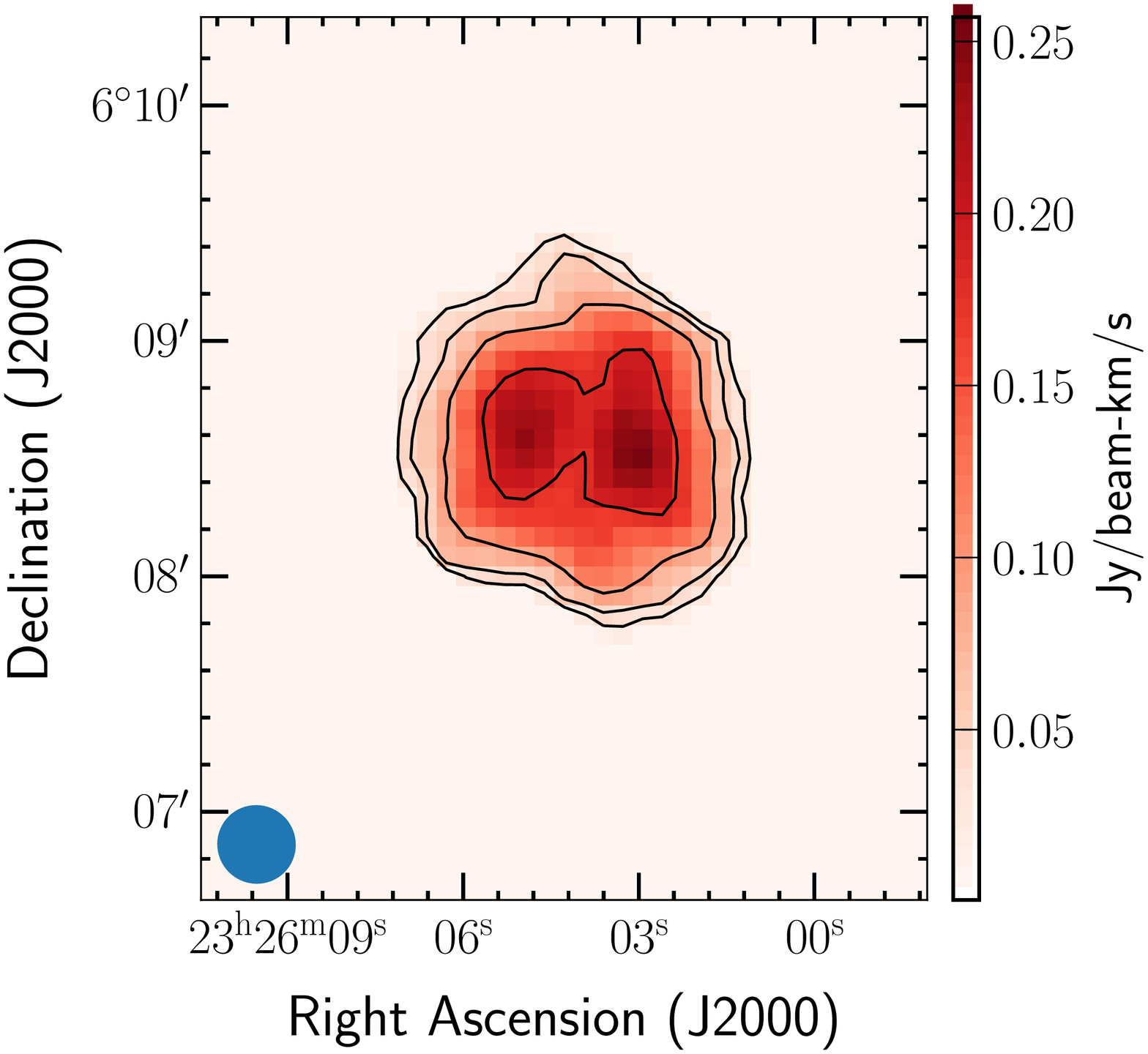}	
		\hspace{0.3cm}	
		\includegraphics[scale=0.23]{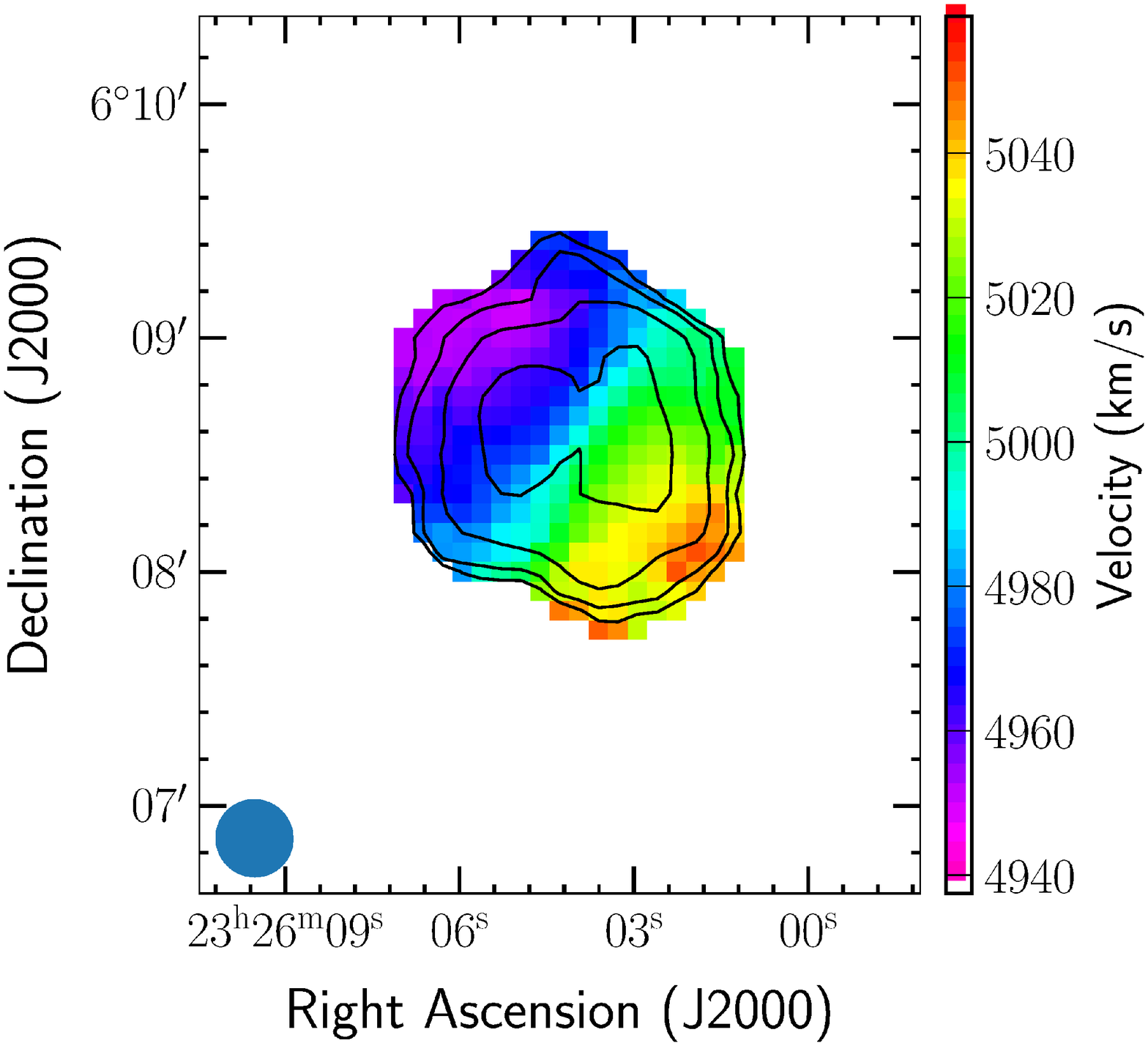}
		\hspace{0.3cm}
		\includegraphics[scale=0.23]{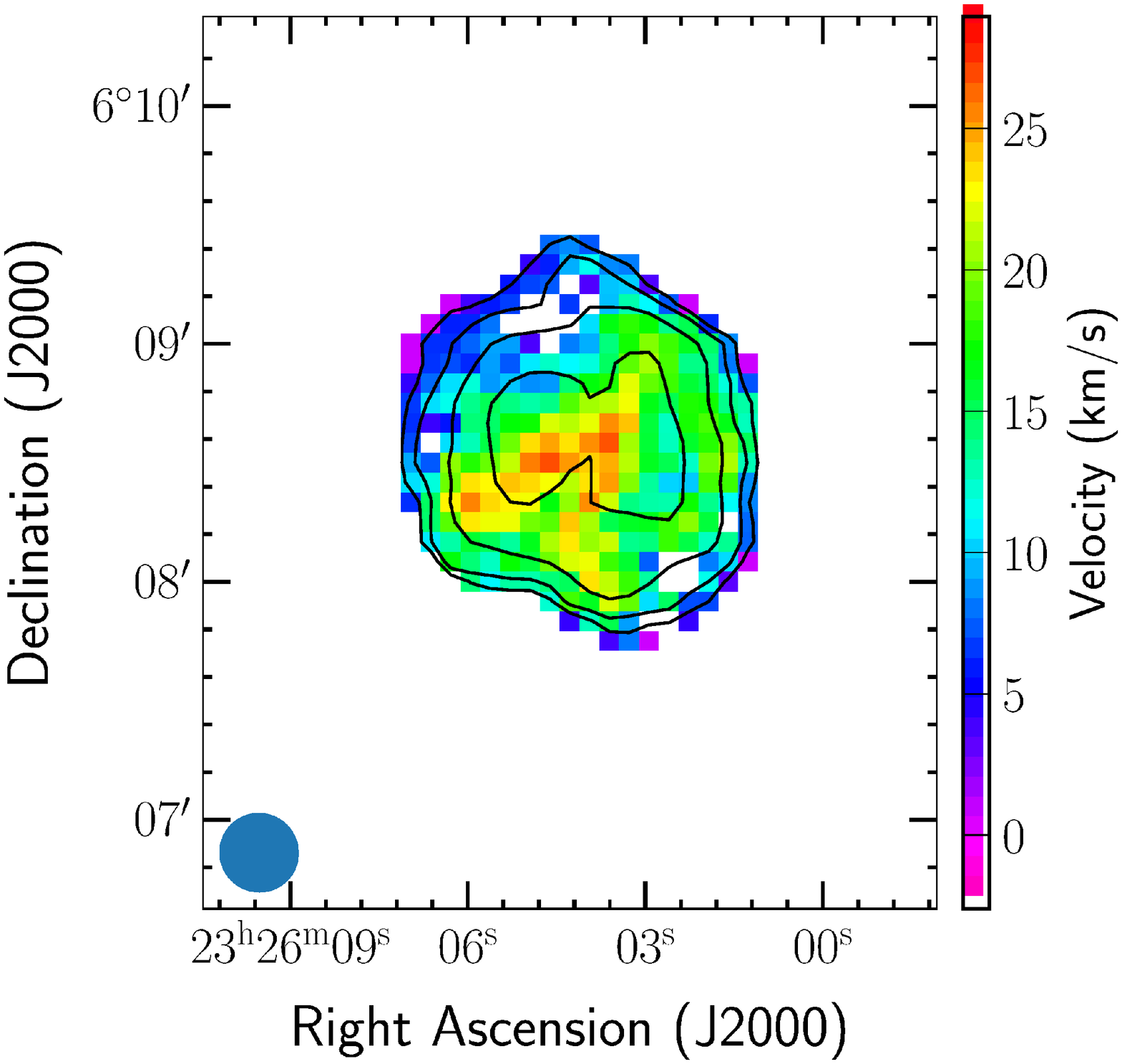}}
	\hbox{	
		\includegraphics[scale=0.23]{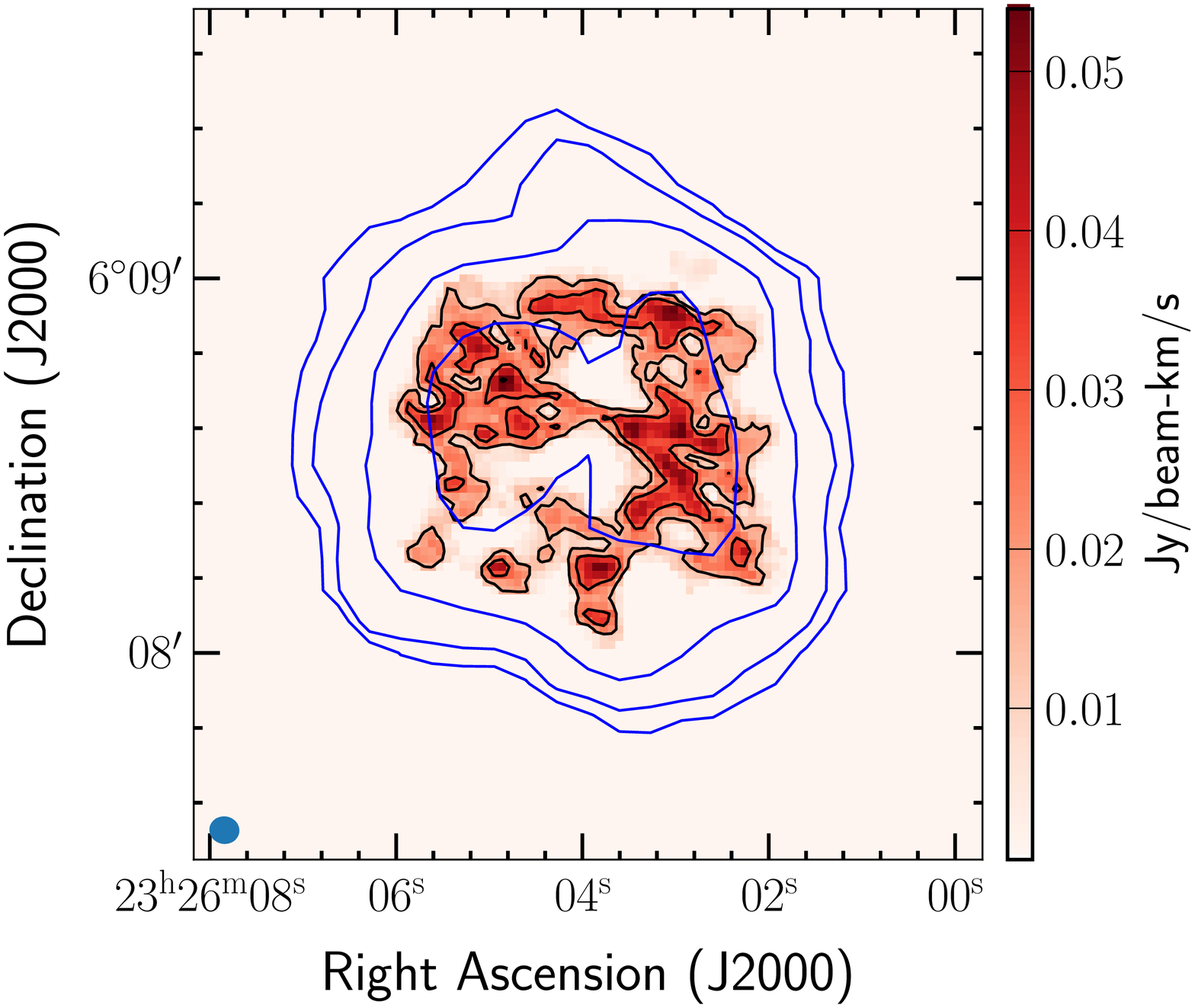}	
		\includegraphics[scale=0.23]{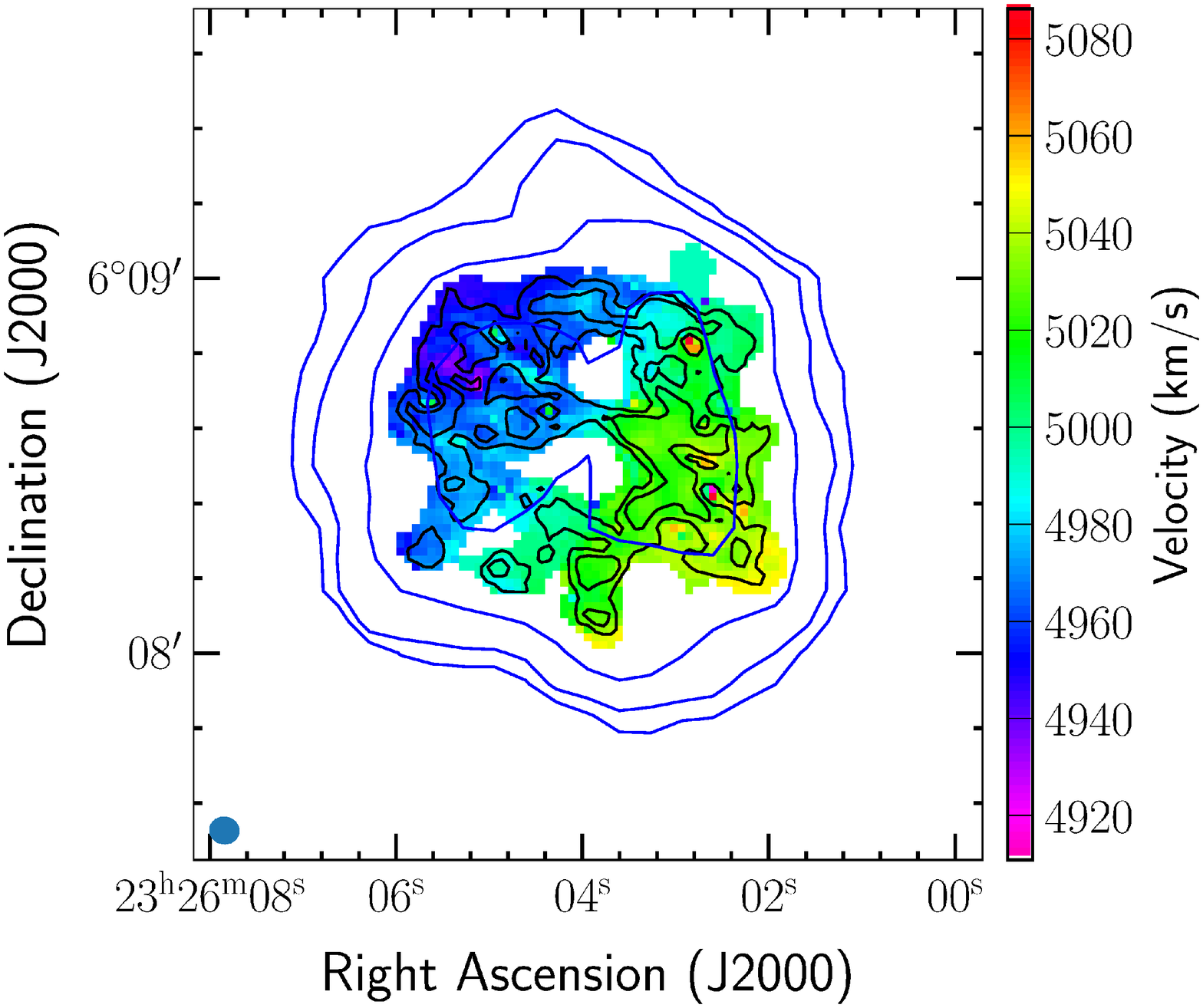}	
		\includegraphics[scale=0.23]{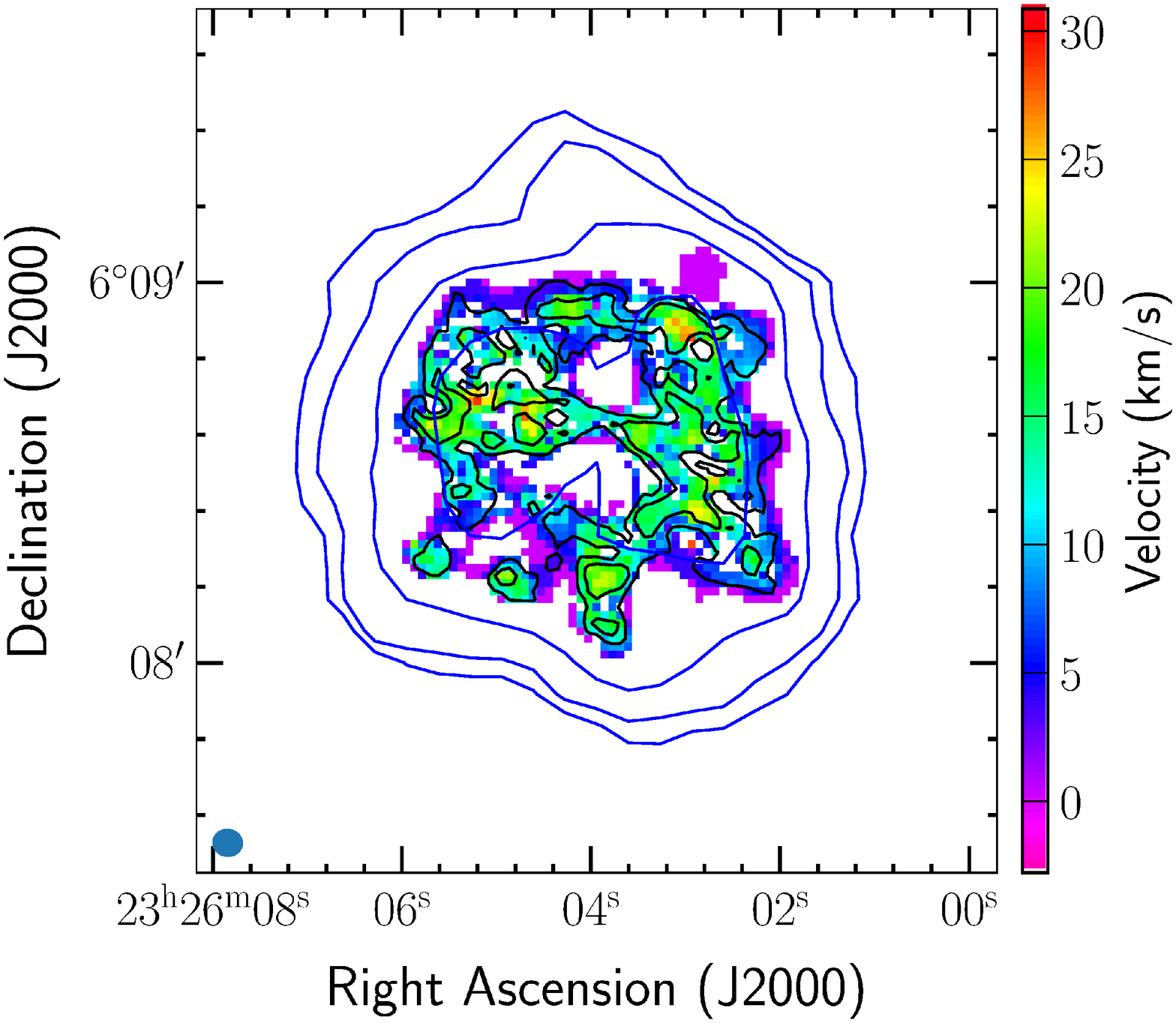}	}	   	
	\caption{
	\textbf{Top panel:}	H{\sc i} integrated flux density contours for W2326+0608 overlaid on moment zero, moment one and moment two maps. The contours are at (1, 2, 4, 8) $\times$ 22.2 mJy beam$^{-1}$ km s$^{-1}$  which correspond to H{\sc i} column densities (1, 2, 4, 8) $\times$ 5.9 $\times$ 10$^{19}$ cm$^{-2}$.  Synthesized beam size is 20.8\arcsec $\times$ 19.9 \arcsec, P.A. 39.0$^{\circ}$ 
	\textbf{Bottom panel:} Higher resolution H{\sc i} integrated flux density contours overlaid on moment zero, moment one and moment two maps. The contours are at  (1, 2, 4) $\times$ 14.8 mJy beam$^{-1}$ km s$^{-1}$  which correspond to  H{\sc i} column densities (1, 2, 4) $\times$ 7.7 $\times$ 10$^{20}$cm$^{-2}$ for a synthesized beam of 4.8\arcsec $\times$  4.4\arcsec, P.A. 81.5$^{\circ}$. Contours from lower resolution in blue colour are also overlaid for comparison.
	}
	\label{fig9}
\end{figure*}

The H{\sc i} gas distribution and kinematics of the BCDGs, W1016+37 and W2326+0608 are shown in Figs.~\ref{fig1}-\ref{fig9} using moment maps and H{\sc i} profiles. The contours in different moment maps are integrated flux densities where the first contours represent 3-$\sigma$ detection limits for a boxcar profile  with a line width $\sim$ 21 km s$^{-1}$. The H{\sc i} channel maps are compiled in the Appendix (Figs.~\ref{fig10}-\ref{fig13} ). The H{\sc i} column densities for the optically thin gas can be estimated from the integrated flux densities~\citep{1978ppim.book.....S} 
\begin{equation}
\frac{N_{\rm H{\sc I}}}{{\rm cm}^{-2}} =  \frac{1.823 \times 10^{18} \times 606}{\theta_{\rm major} \theta_{\rm minor}({\rm arcsec}^{2})}  \int{\frac{S(v)}{\rm mJy}}{\frac{{\rm d}v}{\rm km/s}}
\end{equation} 
Table \ref{sourcecharHI} lists  values for H{\sc i} gas properties derived from different profiles. We  used the Arecibo IDL procedure {\tt mbmeasure}\footnote{\url{http://www.naic.edu/~rminchin/idl/mbmeasure.pro}} to derive the line parameters such as H{\sc i} velocity, Full Width Half Maximum (FWHM), peak flux, {\it rms} noise from line-free channels, integrated flux and H{\sc i}  mass listed in this table. H{\sc i} velocity is the mean of velocities at FWHM. Errors on the H{\sc i} velocity, FWHM, peak flux and integrated flux are calculated using the method of \cite{2004AJ....128...16K}. The H{\sc i}  masses  from H{\sc i} profiles are calculated using the equation~\citep{1975gaun.book..309R}
\begin{equation}
M_{\rm HI} \sim \left(2.36 \times 10^{5}\,{\rm M}_{\odot}\right) \left(\frac{D_{\rm L}}{\mathrm{Mpc}}\right)^{2} \int{\frac{S(v)}{\rm Jy}}{\frac{{\rm d}v}{\rm km/s}}  
\end{equation}  
where $D_{\rm L}$ is luminosity distance and $S(v)$ is the H{\sc i} profile. Luminosity distance has been estimated as  $\sim v_{r}/\mathrm{H_{0}}$ ($v_{r}$ is optical rest velocity). Apart from the error on flux measurement from calibration which is about 5\% to 7\%, the uncertainty of the H{\sc i} mass estimate has a contribution from other factors, mainly the luminosity distance uncertainty which depends on different cosmological parameters. Therefore it is difficult to produce an accurate number for the uncertainty. In the following subsections, we describe H{\sc i} images and profiles for the two sources separately. 

\subsection{W1016+3754}
\label{4.1}  
Low resolution H{\sc i} images ($\sim$ 20\arcsec,  57\arcsec ), as well as high resolution ( $\sim$ 3\arcsec, 10\arcsec ) images, are used to understand the H{\sc i} distribution and kinematics towards the BCDG and its neighbourhood. While lower resolution images are sensitive to diffuse low column density gas ($\sim$10$^{19}$ cm$^{-2}$), higher resolution images provide details of high column density H{\sc i} gas ($\gtrsim$10$^{20}$ cm$^{-2}$).  The H{\sc i} total intensity map with synthesized beam 58.7\arcsec $\times$ 57.0\arcsec, P.A. $-$18.2$^{\circ}$ overlaid on r-band Digitized Sky Survey (DSS) optical image (Fig.~\ref{fig1}) shows a gas-rich galaxy UGC 5540 towards the south at $\sim$38.4 kpc in projection from BCDG W1016+37. In addition to this, we detected another blue compact starburst galaxy, LEDA 2108569 \citep{2015ApJS..217...27A}, at an angular distance of $\sim 6.9'$ towards the south-west from UGC 5540 and redshifted by $\sim$ 100 km s$^{-1}$. \cite{2020MNRAS.498.4745J} reported the presence of  a H{\sc i} cloud towards the east of W1016+3754 in the low resolution ($\sim$ 1\arcmin) image. However, we didn't find evidence of such a cloud from deeper GMRT observations in H{\sc i} map of similar resolution (see the H{\sc i} moment map in Fig.~\ref{fig1} and channel map in Fig.~\ref{fig10}).   

Global H{\sc i} profiles of the BCDG W1016+37, UGC 5540 and LEDA 2108569, extracted from the H{\sc i} cube of the same resolution as these images, are shown in Fig.~\ref{fig2}. Table~\ref{sourcecharHI} lists H{\sc i} profile parameters and log H{\sc i} masses for these sources. The {\it rms} noise from line-free channels in these profiles is 0.9-2.0 mJy per 6.9 km s$^{-1}$. In order to make a comparison, H{\sc i} profile for W1016+3754  from the Arecibo observations (Chandola et al. submitted) is also shown. It shows that H{\sc i} profiles from GMRT and Arecibo observations are similar. Integrated flux for W1016+3754 from the GMRT profile is 0.39$\pm$0.06 Jy km s$^{-1}$ which is similar to integrated flux, 0.49$\pm$0.06 Jy km s$^{-1}$, estimated from the Arecibo profile.  
The  $M_{\rm HI}$  derived from the global H{\sc i} profile of the  BCDG W1016+3754 is $\sim$10$^{7.4}$ M$_{\odot}$ which is similar to the estimated $\sim$10$^{7.5}$ M$_{\odot}$ from our Arecibo observations (Chandola et al. submitted) and similar to an earlier expectation  $M_{\rm HI}$=10$^{7.9}$ M$_{\odot}$ using Nancay Radio Telescope by~\citet{2007A&A...464..859P}. UGC 5540 and LEDA 2108569 have  $M_{\rm HI}$ values $\sim$ 10$^{8.5}$ M$_{\odot}$ and $\sim$ 10$^{7.6}$ M$_{\odot}$ respectively. 

In Fig.~\ref{fig1} bottom left panel, a close-up H{\sc i} map of W1016+3754 at spatial resolution 22.9\arcsec $\times$ 19.6\arcsec, P.A. 24.3$^{\circ}$ overlaid on Sloan Digital Sky Survey (SDSS) r, g, i-band 3 colour image shows that the outer H{\sc i} morphology has somewhat C-shape. The size of H{\sc i} cloud in this image is $\sim$ 1.1\arcmin which corresponds to a linear projected size $\sim$5.3 kpc. However, this C-shape is not seen in the unresolved 20 \arcsec resolution map by \cite{2020MNRAS.498.4745J}.

In Fig.~\ref{fig1} bottom right panel, we also show higher resolution  (10.3\arcsec $\times$ 9.2\arcsec, P.A. 52.7$^{\circ}$) total intensity contours of W1016+37 overlaid on optical SDSS image. The first contour in this image is 18.0 mJy beam$^{-1}$ km s$^{-1}$ which corresponds to H{\sc i} column density of $\sim$ 2.1$\times$ 10$^{20}$ cm$^{-2}$. The total size of H{\sc i} region in this image is $\sim$42\arcsec (projected linear size $\sim$ 3.4 kpc) which is $\sim$4 times as compared to its SDSS optical size r-band size $<$ 10 \arcsec. In Fig.~\ref{fig3}, the moment-zero map shows that there  is also H{\sc i} gas with a density $\sim$10$^{20}$ cm$^{-2}$ towards the north-west at  $\sim$20.6\arcsec (projected linear distance $\sim 1.7\,{\rm kpc}$) distance from the peak column density region which doesn't coincide with any optical feature. This feature is not visible in the high resolution $\sim$8\arcsec map  by \cite{2020MNRAS.498.4745J}. Moment-1 map shows that the gas in this region is redshifted by $\sim$ 30 km s$^{-1}$ relative to the peak H{\sc i} density region.

We further investigated kinematics using the Gaussian fits to the integrated H{\sc i} profile (Fig.~\ref{fig5}) extracted from the cube of the same resolution. The profile fits with two  Gaussian components with parameters listed in Table~\ref{gaussfit}. Two components are separated by  43 km s$^{-1}$ in velocities. The lower flux component (peak flux $\sim$ 4 mJy) is at a redshifted velocity  $\sim$1213 km s$^{-1}$ as compared to the higher flux $\sim$10 mJy component at $\sim$1170 km s$^{-1}$. This implies the presence of two components, the compact dense gas and the diffuse  gas. The latter slightly extends to north-west region and redshifted  velocities. This is also visible in the H{\sc i} channel map (see Fig.~\ref{fig11}) where diffuse components are at velocities higher than $1181.7\,{\rm km}\,{\rm s}^{-1}$ up to $1216.4\,{\rm km}\,{\rm s}^{-1}$.

In Fig. \ref{fig6}, we show H{\sc i} image with higher resolution 3.5\arcsec $\times$ 3.2 \arcsec, P.A. 72.4$^{\circ}$ overlaid on \textit{Hubble Space Telescope (HST) WFC3 F606W} image\footnote{Based on observations made with the NASA/ESA Hubble Space Telescope, and obtained from the Hubble Legacy Archive, which is a collaboration between the Space Telescope Science Institute (STScI/NASA), the Space Telescope European Coordinating Facility (ST-ECF/ESAC/ESA) and the Canadian Astronomy Data Centre (CADC/NRC/CSA).}. The high column density H{\sc i} gas ($\sim$10$^{21}$ cm$^{-2}$) seems to have a tadpole-like structure similar to seen in SDSS/HST optical images. The gas with peak column density, N(H{\sc i}) $\sim$7.6 $\times$10$^{21}$ cm$^{-2}$, is near the region of higher stellar density. It appears that the H{\sc i} peak position might be shifted toward the east by 2.5 \arcsec or a projected distance of 200 parsecs. However, we refrain from making any  claim on this shift as it is  smaller than the resolution of the H{\sc i} image.

\subsection{W2326+0608}
\label{4.2}
Low resolution (20.8\arcsec $\times$ 19.6\arcsec, P.A. 28.8$^{\circ}$) image  (Fig.~\ref{fig1a}, which is sensitive enough to detect H{\sc i} gas with column density $\gtrsim$ 5.9 $\times$ 10$^{19}$ cm$^{-2}$, shows both W2326+0608 and nearby galaxy SDSS J232603.86+060835.8 share  same H{\sc i} environment. H{\sc i} region in this image extends up to $\sim$1.7\arcmin  which corresponds to projected linear size  $\sim$35 kpc. Integrated H{\sc i} profile for H{\sc i} region in this image is extracted from the same low resolution H{\sc i} cube (Fig.~\ref{fig7} ). The $rms$ noise in this profile is 1.4 mJy per 7.1 km s$^{-1}$. This H{\sc i}  profile is similar to the profile from  Arecibo observations (Chandola et al. submitted) towards this source.
Integrated flux from this GMRT H{\sc i} profile is $1.63 \pm 0.17$ Jy km s$^{-1}$  compared to the integrated flux $1.67\pm 0.13$\,Jy km s$^{-1}$ from the Arecibo H{\sc i} profile.  H{\sc i} mass derived from this profile is $10^{9.3}\,{\rm M}_{\odot}$.   The double horn that is seen in both Arecibo and GMRT integrated H{\sc i} line profile (Fig.~\ref{fig7}) indicate the rotating motion of the gas.

We also mapped H{\sc i} towards W2326+0608 with a resolution of 4.8\arcsec $\times$ 4.4\arcsec, P.A. 81.46$^{\circ}$ which was sufficient to detect high column density H{\sc i} gas of N(H{\sc i}) $\gtrsim$ 7.7$\times$10$^{20}$ cm$^{-1}$ and resolve the H{\sc i} towards the BCDG and neighbouring galaxy (Fig.~\ref{fig8}).  With this high-resolution image, we are able to detect regions of very high H{\sc i} column densities ($\sim$ 10$^{21}$ cm$^{-2}$ ), including one near the BCDG W2326+0608. The size of this H{\sc i} region is around $\sim$ 10\arcsec (projected linear size $\sim$3.4 kpc), and is shifted by $\sim$4.6 \arcsec (projected distance $\sim$1.6 kpc) towards east from W2326+0608. 
We also show the H{\sc i} line profile (Fig.~\ref{fig8} right panel) towards this high H{\sc i} density region extracted from a circle marked in green in the left panel. The $rms$ noise in this profile is 0.4 mJy per 7.1 km s$^{-1}$.  The total integrated flux from this profile is 0.05$\pm$0.03 Jy km s$^{-1}$ which corresponds to the H{\sc i} mass of 10$^{7.8}$ M$_{\odot}$ (Table~\ref{sourcecharHI}).  Most of the high column density H{\sc i} gas regions are not aligned along the major axis of edge-on galaxy J232603.86+060835.8 and are found to be at outskirts. Except for the gas which is associated with the BCDG W2326+0608, most of the gas at the outskirts is with no optical/UV counterpart.  The moment one maps in Fig.~\ref{fig9}  show a velocity gradient for the gas along the major axis of J232603.86+060835.8 indicating the rotating motion, though slightly tilted.

\section{Discussion}
\label{sec5}

\subsection{What triggered the starburst in W1016+37 and W2326+0608 ?}  
Starburst in low mass galaxies like BCDGs could be triggered due to different external \citep{1987ApJ...322L..59S,1988A&A...201...37N,2001A&A...371..806N, 2008MNRAS.388L..10B,lopezsanchez2010A&A...521A..63L,  2014MNRAS.445.1694L,2015ApJ...802...82F} or internal mechanisms \citep{2004AJ....128.2170H,2012ApJ...747..105E}.  Among the external mechanisms, there could be merger/interaction with the neighbouring galaxies \citep{mendez2000A&A...359..493M} or accretion of intergalactic pristine gas \citep{sanchezalmeida2015ApJ...810L..15S,2015ApJ...802...82F}.  \cite{2015ApJ...802...82F} found most of the extremely low metallicity systems in their sample are located in low-density environments. They suggested that star formation is fuelled by the accretion of metal-poor gas similar to cold-gas inflows at high redshift. The evidences for merger or interaction are very faint in optical images and could be seen in radio H{\sc i} images \citep{lopezsanchez2008A&A...491..131L,lopezsanchez2010A&A...521A..63L,2014MNRAS.445.1694L}. According to \citet{2014MNRAS.445.1694L}, young starburst dwarf galaxies  have more outer asymmetries in H{\sc i} images and  external mechanisms like interactions or infall of gas are responsible for that.  
The asymmetry in W1016+3754 GMRT 10.3 \arcsec $\times$ 9.2\arcsec, P.A. 52.7$^{\circ}$ H{\sc i} map and profile is consistent with this. The presence of red-shifted diffuse gas  near this BCDG hints at the scenario that the infalling cold gas fuels the starburst activity in BCDGs.
Since the  projected distance ($\sim$38.4 kpc) of W1016+3754 from UGC 5540 is close enough (so as the star-burst dwarf irregular LEDA 2108569) to cause gravitational perturbation \citep{2008gady.book.....B, 2015ApJ...802...82F},  this  infall of gas in W1016+3754 might be triggered by UGC 5540. Although  no direct evidence of tidal interaction  like tidal filaments is visible in H{\sc i} image possibly due to the sensitivity limit of our observation.  W2326+0608 is located in the H{\sc i} gas-rich environment of nearby galaxy SDSS J232603.86+060835.8 and they are tidally interacting. Due to this tidal interaction, there might be a collapse or infall of gas which has triggered the starburst in W2326+0608.  
Though we do not have enough spatial resolution to look into the details of W2326+0608 itself, the asymmetric H{\sc i} profile hints at this scenario.
  
\subsection{H{\sc i} gas properties and star formation}
Several previous studies have studied the link between H{\sc i} gas properties and star formation in dwarf galaxies. Galaxies with higher star formation rate surface densities were found to have higher H{\sc i} surface densities \citep{2014A&A...566A..71L}.  \cite{1987NASCP2466..263S} in their studies of irregular dwarf galaxies estimated critical H{\sc i} column density for massive star formation to be $\sim$ 10$^{21}$ cm$^{-2}$. However, in some  studies, star-formation at sub-critical H{\sc i} column densities have been also reported \citep{2016AJ....152..202C, 2016ApJ...832...85T}. Our study towards W1016+37 and W2326+0608 is consistent with  star formation taking place in the region of high H{\sc i} column densities of $\sim$10$^{21}$ cm$^{-2}$.  The peak H{\sc i} column densities for W1016+3754 and W2326+0608 are $(7.6\pm 1.6)\times 10^{21}\,{\rm cm}^{-2}$ and $(2.7\pm 0.6)\times 10^{21}\,{\rm cm}^{-2}$ respectively. There appears to be some difference between the position of  the peak of the two parameters in both sources.  The physical reasons behind these differences can be related to the star formation history or tidal interactions  \citep{2016ApJ...832...85T}. In W1016+3754 high-resolution image (Fig. \ref{fig6}) the shift of $\sim$ 200 pc  may  be from shock waves due to Supernovae explosion or star-burst, evidence for which comes from the presence of high ionization lines \citep{2012MNRAS.427.1229I}. Similarly, in the case of W2326+0608, it is possible strong stellar winds from recent starburst (SFR $\sim$ 0.3 M$_{\odot}$ yr$^{-1}$) or tidal interactions with SDSS J232603.86+060835.8 have caused the shift of $\sim$ 1.6 kpc as it could have strong gravitational effect from a projected distance of $\sim$8 kpc.  

We further analyse the gas and star-formation conditions in these two galaxies using the model of \cite{2004A&A...421..555H}. Initial gas densities, ambient pressure due to shock waves and dust enrichment from Supernovae Type {\sc ii}   are crucial factors for the formation of dense and compact super star clusters (SSC) seen in MIR bright BCDGs as these can trigger the runaway star formation \citep{2002AJ....123.1454B, 2004A&A...421..555H}. For a pressure-bound and self-gravity driven isothermal spherical star-forming region, the critical radius is $\sim$ 50 pc below which it would become unstable, collapse and turn into SSC \citep{1955ZA.....36..222E,1956MNRAS.116..351B, 2004A&A...421..555H}. We could only determine the radius of the star-forming region ($r_{\rm SF}$) for W1016+3754 from its high resolution \textit{HST WFCS F606W} image (Fig.~\ref{fig6}). The image shows mainly three super star cluster regions of average $r_{\rm SF}$ $\sim$0.5\arcsec $\sim$ 40 pc. Since we don't have a high resolution image for W2326+0608, we assume an upper limit on its $r_{\rm SF}< 50\,{\rm pc}$. If we assume all of the total H{\sc i} gas mass of these galaxies is associated with the star-formation regions, then according to eq.~(7) in \citet{2004A&A...421..555H}, the gas number densities for W1016+3754 and W2326+0608 are $\sim 3.9\times 10^{3}\,{\rm cm}^{-3}$ and $\gtrsim$ 5$\times 10^{3}\,{\rm cm}^{-3}$  respectively. These gas number densities correspond to a free-fall timescale ($t_{\rm ff}$) $\sim 1.6\,{\rm Myr}$ and $\lesssim\, 1.4\,{\rm Myr}$ according to {eq.~(4) in \cite{2004A&A...421..555H}. These time scales are similar to the starburst age of $\sim$2 Myr estimated from EQW(H${\beta}$) for W1016+3754. At the current rate of star formation, $\sim$$0.04\,{\rm M}_{\odot}\,{\rm yr}^{-1}$ in W1016+3754 and $\sim$$0.3\,{\rm M}_{\odot}\,{\rm yr}^{-1}$  in  W2326+0608, the gas consumption time scale is  $\sim$$6.3\times$10$^{8}$ yrs for W1016+3754 and $\sim$$2.1\times$10$^{8}$ yrs for W2326+0608. According to \cite{2004A&A...421..555H} model, SNe II  starts dust enrichment in active star-forming regions on the time scales of $\sim$3 Myr after the onset of star-formation. Initially, the dust temperature ($T_{\rm dust}$) is a few 100 K and cools down rapidly. Dust temperature  also has a dependence on the dust grain size($d_{\rm dust}$) as $d_{\rm dust}^{-1/6}$.  Hence the presence of small grain size dust from SNe {\sc ii} in W1016+3754 and W2326+0608, and their age of starburst and freefall time scale explains the bright red MIR colours. 
\begin{figure}
	\centering
	\vbox{

			\includegraphics[scale=0.45]{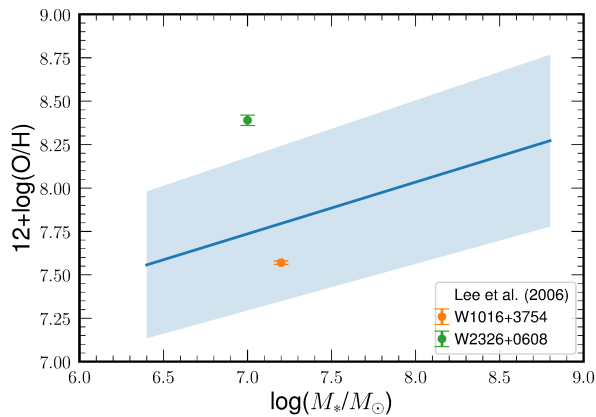}
	}    				
	\caption{Oxygen abundance vs stellar mass plot shows the position of two galaxies in orange and green colours for W1016+3754 and W2326+0608 respectively. Blue solid line represents the stellar mass-metallicity relation by \protect\cite{lee2006ApJ...647..970L} for dwarf galaxies. Blue shaded region shows the dispersion in mass-metallicity relation by \protect\cite{lee2006ApJ...647..970L}.}
	\label{figmz}
\end{figure}  
\subsection{The position of two galaxies in mass-metallicity diagram}
In Fig.~\ref{figmz}, we show the position of two galaxies in the mass-metallicity diagram. The metallicity of W1016+3754 for its stellar mass is below the mass-metallicity relation given by \cite{lee2006ApJ...647..970L} for dwarf galaxies.
Since the starburst is fresh, and metal-free gas is still accumulating, the metallicity at this stage is low for W1016+3754.  This is consistent with the analysis of~\citet{2018MNRAS.477..392L} where they found the lowest oxygen abundance in the regions of disturbed neutral and ionized gas kinematics in the starburst galaxy UGC 461. \cite{sanchezalmeida2015ApJ...810L..15S} find inhomogeneous metallicity distribution in their sample of dwarf galaxies with starburst regions having lowest metallicity suggesting infall of metal-poor gas triggering the starburst.
This is also consistent with the scenario in NGC 5253 where \cite{lopezsanchez2012MNRAS.419.1051L} found direct evidence of infalling metal-poor H{\sc i} gas.

However, in the case of W2326+0608, we find that despite having similar stellar mass and higher sSFR compared to W1016+3754, the metallicity is higher by nearly an order of magnitude. In the fundamental metallicity relation \citep{mannucci2010MNRAS.408.2115M}, those having higher sSFR have lower metallicities which W2326+0608 doesn't appear to follow.  The common H{\sc i} environment with the galaxy SDSS J232603.86+060835.8 suggests higher metallicity of W2326+0608 is due to its proximity and metal transport from neighbouring galaxy \citep{2009ApJ...705..723C,2015MNRAS.450.2367R}. In future, many such systems with young starburst  will be discovered with Commensal Radio Astronomy FasT Survey \citep[CRAFTS;][]{2018IMMag..19..112L,2021MNRAS.500.1741Z} using the Five-hundred-meter Aperture Spherical radio Telescope (FAST).

\section{Conclusion}
\label{sec6}
In this section, we summarize our findings from the GMRT H{\sc i} study of the bright-MIR BCDGs W1016+3754 and W2326+0608.
\begin{itemize}
	\item Our data suggest that the infall of diffuse cold gas has triggered the star formation in both BCDGs. The origin of the cold gas could be the consequence of gas expelled by interactions between neighbouring galaxies, or just gas accretion from the intergalactic medium.

	\item Star formation in both the galaxies takes place in the regions of H{\sc i} gas column densities greater than 10$^{21}\,{\rm cm}^{-2}$.   

	\item 
	In the case of W1016+3754, we find evidences of infall of gas as the triggering mechanism of the recent starburst. The metallicity of the gas is lower than given by the mass-metallicity relation for its stellar mass, suggesting that the infalling gas is metal-poor.

	\item 
	Although W2326+0608 has a higher specific star-formation rate and similar stellar mass as W1016+3754, the gas has a much higher metallicity by nearly an order of magnitude above mass-metallicity relation, suggesting that the infalling gas is metal-rich, perhaps as a consequence of tidal interaction with the neighbouring galaxy SDSS J232603.86+060835.8.

\item In future, many low metallicity blue compact dwarf galaxies with recently triggered star formation will be discovered with the CRAFTS extragalactic H{\sc i} survey. 	

\end{itemize} 
\section*{Data availability}
GMRT data can be downloaded from the GMRT data archive \url{https://naps.ncra.tifr.res.in/goa/data/search} using the project codes 26\_060 and 28\_091.
\section*{Acknowledgements}
We thank the anonymous reviewer for the useful comments which  helped to significantly improve the paper.
We thank the staff of the GMRT that made these observations possible. GMRT is run by the National Centre for Radio Astrophysics of the Tata Institute of Fundamental Research. We used AIPS for data reduction which is produced and maintained by the National Radio Astronomy Observatory, a facility of the National Science Foundation operated under cooperative agreement by Associated Universities, Inc. CWT thanks D. Stern for attempting to get the supporting optical spectra for this paper. YC thanks Narendranath Patra for the useful discussion on data reduction. 

In this work, YC is sponsored by the Chinese Academy of Sciences Visiting Fellowship for Researchers from Developing Countries, Grant No. 2013FFJB0009. YC also thanks Center for Astronomical Mega-Science, CAS, for FAST distinguished young researcher fellowship (19-FAST-02) and China ministry of science and technology (MOST) for the grant no. QNJ2021061003L. YC also acknowledges support from National Natural Science Foundation of China (NSFC) Grant No. 11550110181 and 12050410259. 
CWT was supported by a grant from the NSFC (No. 12041302).
 YZM is supported by the National Research Foundation of South Africa under grant No. 120385 and No. 120378, NITheCS program ``New Insights into Astrophysics and Cosmology with Theoretical Models confronting Observational Data'', and National Natural Science Foundation of China with project 12047503. 

This publication makes use of data products from the \textit{Wide-Field Infrared Survey Explorer}, which is a joint project of the University of California, Los Angeles, and the Jet Propulsion Laboratory, California Institute of Technology, funded by the National Aeronautics and Space Administration. The Arecibo Observatory is operated by SRI International under a cooperative agreement with the National Science Foundation (AST-1100968), and in alliance with Ana G. M\'endez-Universidad Metropolitana, and the Universities Space
Research Association. 

This research has made use of the NASA/ IPAC Infrared Science Archive, which is operated by the Jet Propulsion Laboratory, California Institute of Technology, under contract with the National Aeronautics and Space Administration.  This research has made use of the VizieR catalogue access tool, CDS, Strasbourg, France. The original description of the VizieR service was published in A\&AS 143, 23. 

This research has made use of Digitized Sky Survey products. The Digitized Sky Survey was produced at the Space Telescope Science Institute under U.S. Government grant NAG W-2166. The images of these surveys are based on photographic data obtained using the Oschin Schmidt Telescope on Palomar Mountain and the UK Schmidt Telescope. The plates were processed into the present compressed digital form with the permission of these institutions. 

This work also makes use of Sloan Digital Sky Survey (SDSS)-III. Funding for SDSS-III has been provided by the Alfred P. Sloan Foundation, the Participating Institutions, the National Science Foundation
and the US Department of Energy Office of Science. The SDSS-III web site is http://www.sdss3.org/. SDSS-III is managed by the Astrophysical Research Consortium for the Participating Institutions of the SDSS-III Collaboration including the University of Arizona, the Brazilian Participation Group, Brookhaven National Laboratory, Carnegie Mellon University, University of Florida, the French Participation Group, the German Participation Group, Harvard University, the Instituto de Astrofisica de Canarias, the Michi-
gan State/Notre Dame/JINA Participation Group, Johns Hopkins University, Lawrence Berkeley National Laboratory, Max Planck Institute for Astrophysics, Max Planck Institute for Extraterrestrial Physics, New Mexico State University, New York University, Ohio State University, Pennsylvania State University, University of Portsmouth, Princeton University, the Spanish Participation Group, University of Tokyo, University of Utah, Vanderbilt University, University of Virginia, University of Washington and Yale University. Some of the data presented in this paper were obtained from the Mikulski Archive for Space Telescopes (MAST). STScI is operated by the Association of Universities for Research in Astronomy, Inc., under NASA contract NAS5-26555. Support for MAST for non-HST data is provided by the NASA Office of Space Science via grant NNX09AF08G and by other grants and contracts. This work has also used different Python packages e.g. NUMPY, ASTROPY, APLPY,
SCIPY and MATPLOTLIB. We thank numerous contributors to these packages. 


\appendix
\section{}
\clearpage
\begin{figure*}
	\centering
	\hbox{ 				
		\includegraphics[scale=0.9]{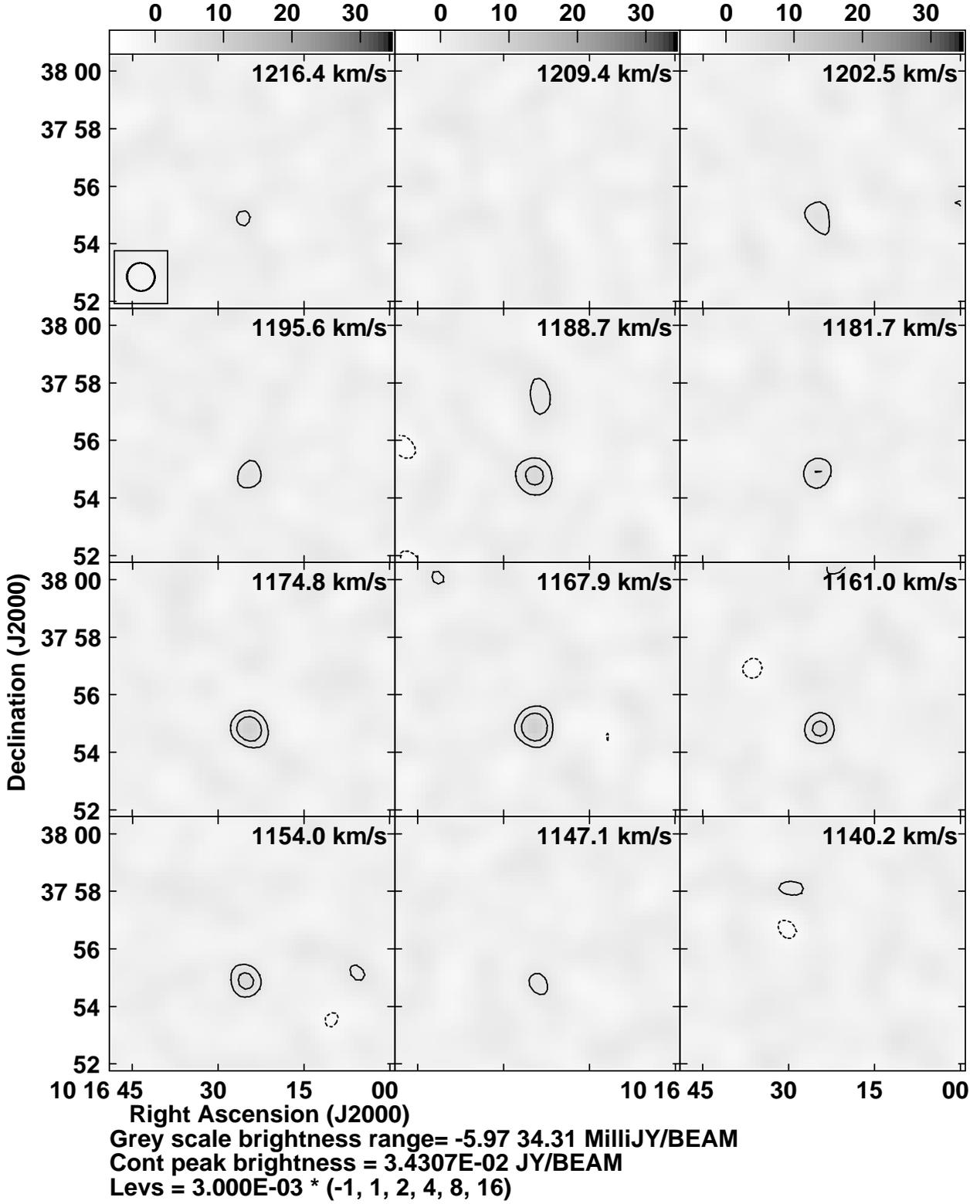}		
	   }
	\caption{
		H{\sc i} flux density contours overlaid on grey scale channel maps for W1016+3754. The contours are at (-1, 1, 2, 4,8,16) $\times$ 3  mJy beam$^{-1}$  which correspond to  H{\sc i} column densities (-1, 1, 2, 4, 8, 16) $\times$6.8$\times$ 10$^{18}$cm$^{-2}$ for a synthesized beam of 58.7\arcsec $\times$ 57.0\arcsec, P.A. 20.4$^{\circ}$.   		
	}
	\label{fig10}
\end{figure*} 
\begin{figure*}
	\centering
	\hbox{ 				
		\includegraphics[scale=0.9]{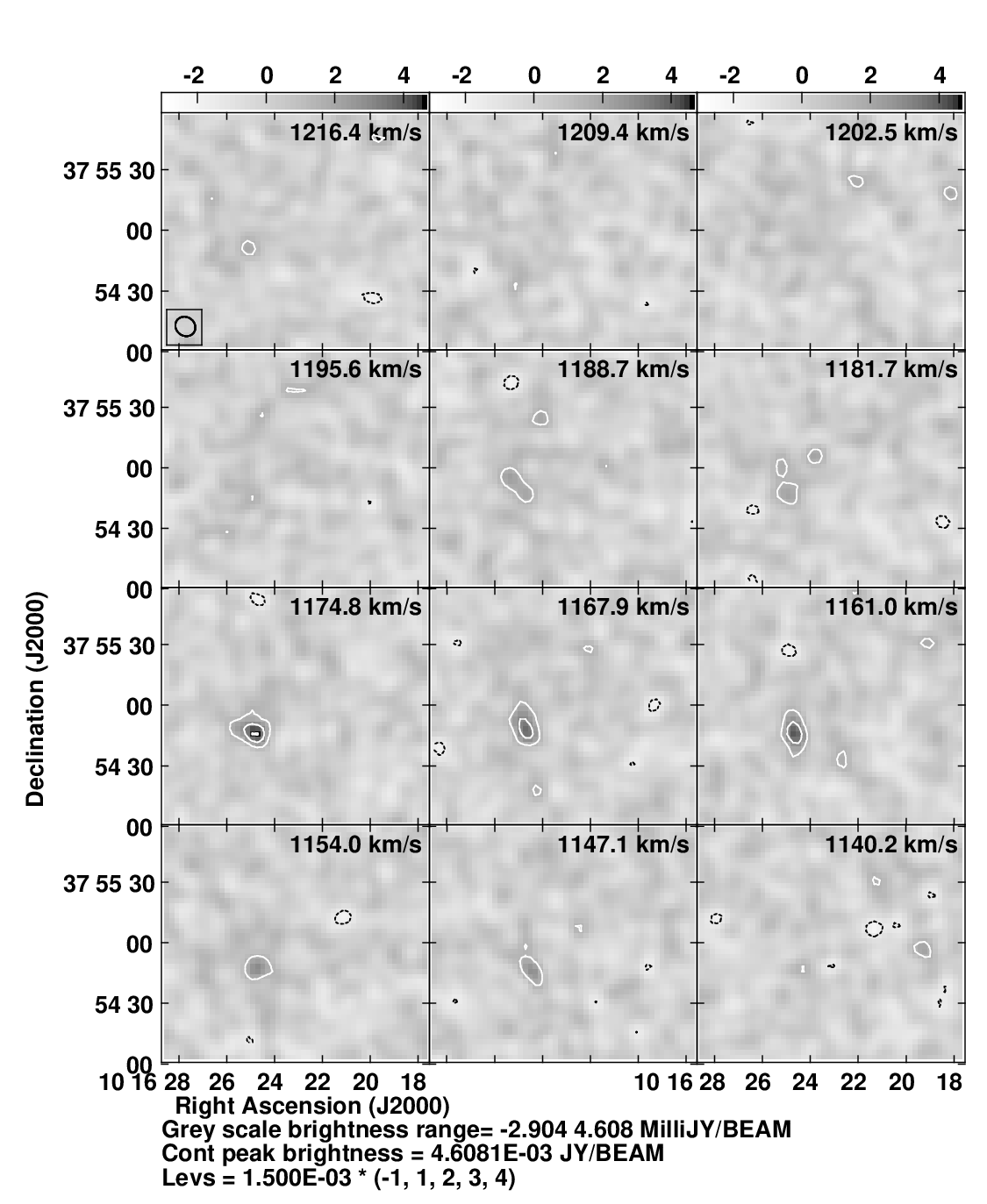}		
	}
	\caption{H{\sc i} flux density contours overlaid on grey scale channel maps for W1016+3754. The contours are at (-1, 1, 2, 3, 4) $\times$1.5 mJy beam$^{-1}$  which correspond to  H{\sc i} column densities (-1, 1, 2, 3,4) $\times$1.20 $\times$ 10$^{20}$cm$^{-2}$ for a synthesized beam of 10.3\arcsec $\times$ 9.2\arcsec, P.A. 52.7$^{\circ}$.	   		
	}
	\label{fig11}
\end{figure*} 

\begin{figure*}
	\centering
	\hbox{ 				
		\includegraphics[scale=0.9]{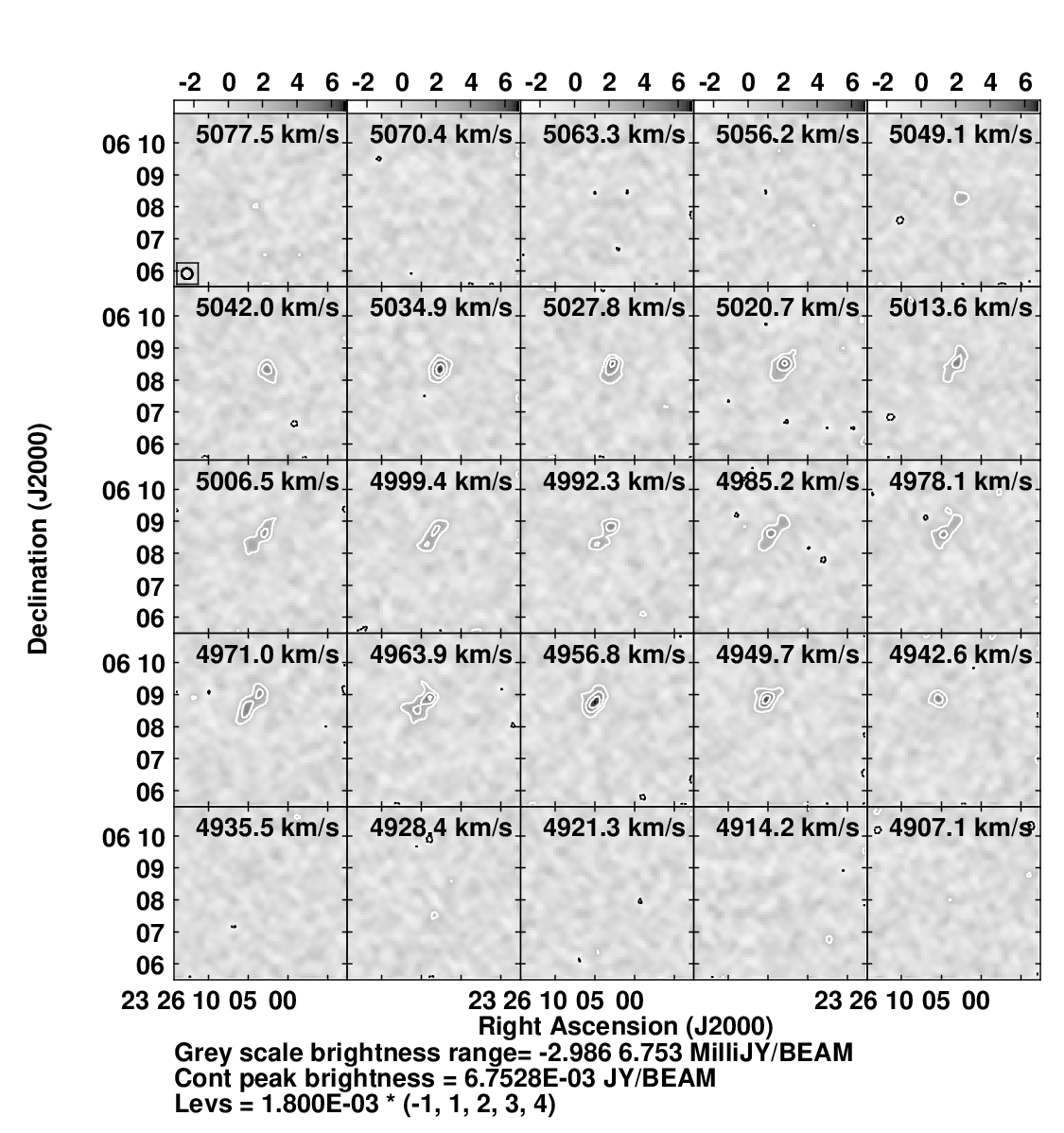}		
	}
	\caption{H{\sc i} flux density contours overlaid on grey scale channel maps for W2326+0608. The contours are at (-1, 1, 2, 3, 4) $\times$1.8 mJy beam$^{-1}$  which correspond to  H{\sc i} column densities (-1, 1, 2, 3,4) $\times$34.1 $\times$ 10$^{18}$cm$^{-2}$ for a synthesized beam of 20.8\arcsec $\times$ 19.9\arcsec, P.A. 39$^{\circ}$.	
	}
	\label{fig12}
\end{figure*} 
\begin{figure*}
	\centering
	\hbox{ 				
		\includegraphics[scale=0.9]{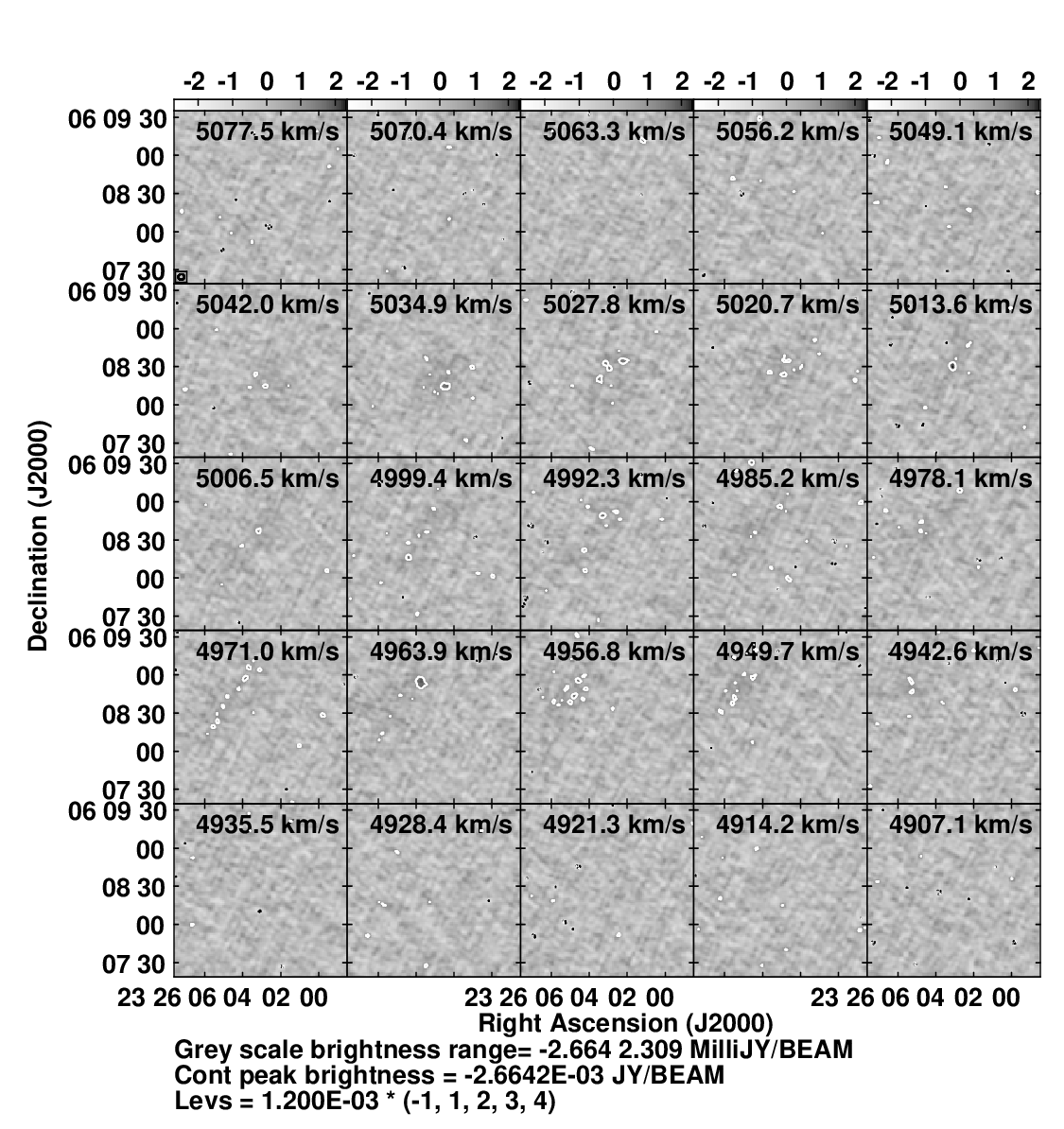}		
	}
	\caption{
H{\sc i} flux density contours overlaid on grey scale channel maps for W2326+0608. The contours are at (-1, 1, 2, 3, 4) $\times$1.2 mJy beam$^{-1}$  which correspond to  H{\sc i} column densities (-1, 1, 2, 3,4) $\times$4.45 $\times$ 10$^{20}$cm$^{-2}$ for a synthesized beam of 4.8\arcsec $\times$ 4.4\arcsec, P.A. 81.46$^{\circ}$.
	}
	\label{fig13}
\end{figure*} 
\end{document}